\documentclass[fleqn,usenatbib]{mnras}

\usepackage{newtxtext,newtxmath}

\usepackage[T1]{fontenc}
\usepackage{ae,aecompl}

\usepackage{graphicx}	
\graphicspath{ {figures/} }
\usepackage{amsmath}	
\usepackage{amssymb}	
\usepackage{subcaption} 
\newcommand{\un}[2]{\,\text{#1}^{#2}} 
\newcommand{\Myr}{\un{Myr}{}}
\newcommand{\kms}{\un{km}{}\un{s}{-1}}

\newcommand{\vm}{\boldsymbol{\mu}}
\newcommand{\vh}{\boldsymbol{h}}
\newcommand{\bS}{\boldsymbol{\Sigma}}
\newcommand{\bZ}{\boldsymbol{Z}}
\newcommand{\vt}{\boldsymbol{\theta}}
\newcommand{\vf}{\mathbf{f}}
\newcommand{\bj}{\mathbf{J}}
\newcommand{\Norm}[3]{\mathcal{N}(#1;#2,#3)} 
\newcommand{\tht}[2]{\theta_{#1,#2}}

\newcommand{\aref}[1]{\hyperref[#1]{Appendix~\ref{#1}}}

\title[Chronostar. Derivation and application]{\texttt{Chronostar}: 
a novel Bayesian method for kinematic age determination. 
I. Derivation and application to the $\beta$ Pictoris 
moving group
}

\author[T. D. Crundall et al.]{Timothy D. Crundall,$^{1,3}$\thanks{E-mail: timothy.crundall@anu.edu.au (ANU)}
Michael J. Ireland,$^{1}$,
Mark R. Krumholz,$^{1,2}$
\newauthor{Christoph Federrath,$^{1,2}$
Maru\v sa \v Zerjal$^{1}$
and Jonah T. Hansen$^{1}$}
\\
$^{1}$Research School of Astronomy and Astrophysics, Australian National University, Canberra 2600, Australia\\
$^{2}$ARC Centre of Excellence for Astronomy in Three Dimensions (ASTRO-3D), Australia\\
$^{3}$I. Physikalisches Institut, Universit{\" a}t zu K{\" o}ln, Z{\" u}lpicher Str. 77, D-50937 K{\" o}ln, Germany
}

\date{Accepted 2019 August 22. Received 2019 August 14; in original form 2019 February 15}

\pubyear{2019}

\begin{document}
\label{firstpage}
\pagerange{\pageref{firstpage}--\pageref{lastpage}}
\maketitle

\begin{abstract}
\textit{Gaia} DR2 provides an unprecedented sample of stars with full 6D phase-space measurements, creating the need for a self-consistent means of discovering and characterising the phase-space overdensities known as \textit{moving groups} or \textit{associations}. Here we present \texttt{Chronostar}, a new Bayesian analysis tool that meets this need. \texttt{Chronostar} uses the Expectation-Maximisation algorithm to remove the circular dependency between association membership lists and fits to their phase-space distributions, making it possible to discover unknown associations within a kinematic data set. It uses forward-modelling of orbits through the Galactic potential to overcome the problem of tracing backward stars whose kinematics have significant observational errors, thereby providing reliable ages. In tests using synthetic data sets with realistic measurement errors and complex initial distributions, \texttt{Chronostar} successfully recovers membership assignments and kinematic ages up to $\approx 100$ Myr. In tests on real stellar kinematic data in the phase-space vicinity of the $\beta$ Pictoris Moving Group, \texttt{Chronostar} successfully rediscovers the association without any human intervention, identifies 15 new likely members, corroborates 43 candidate members, and returns a kinematic age of  $17.8\pm 1.2$\,Myr. In the process we also rediscover the Tucana-Horologium Moving Group, for which we obtain a kinematic age of $36.3^{+1.3}_{-1.4}$\,Myr.
\end{abstract}




\section{Introduction}
With the advent of \textit{Gaia} DR2 \citep{gaia_collaboration_gaia_2018} 
we have access to an all-sky, magnitude complete survey 
that provides full 6D kinematic information for over 
7,000,000 stars. Within this wealth of data reside 
the kinematic fingerprints of star formation events 
in the form of moving groups, stars that were formed 
in close proximity (both spatially and temporally) 
that have since become unbound and are now following 
approximately
ballistic trajectories through the Galaxy.
The development of an accurate and reliable method to infer 
the origin site of a moving group is a critical step 
in using kinematic information to constrain
stellar ages, which in turn would allow calibration 
of model dependent ageing techniques.
Accurate ages are important for many applications. They set the clock for circumstellar disc evolution and planet formation.
Exoplanets are most easily directly imaged when they are young,
so accurate ages enable better target selection for direct imaging campaigns.
Accurate ages are required for calibration of massive stellar evolution models, but are nearly impossible to obtain directly due to these stars' short Kelvin-Helmholtz contraction times; however, they can be age-dated approximately via their association with less massive members of a moving group.

However, current kinematic analysis methods have proven unable to deliver age estimates that are consistent with one another, or with other age estimators.
One common kinematic approach is to estimate a \textit{traceback age} by following the orbits
of group members backwards through time to identify the age at
which they occupied the smallest spatial volume. \citet{ducourant_tw_2014}
employ this technique to obtain a kinematic age for the TW Hydrae
Association (TWA) of $7.5\pm0.7\Myr$. However
\citet{donaldson_new_2016}
obtain a different age of $3.8 \pm
1.1 \Myr$ using the same method, a discrepancy that they attribute to \citeauthor{ducourant_tw_2014} not properly propagating measurement uncertainties. 
\cite{mamajek_age_2014} 
review age estimates for 
the $\beta$ Pictoris Moving Group (henceforth $\beta$PMG) 
and find that traceback 
ages (\citealt{ortega_origin_2002}, \citealt{song_new_2003}, 
\citealt{ortega_new_2004}) 
are $4\sigma$-discrepant
with the combined lithium depletion boundary (LDB) 
and isochronal age of $23 \pm 3 \Myr$; the sole exception 
is the traceback age of $22\pm12\Myr$ determined by 
\citet{makarov_unraveling_2007}, which has such a large 
uncertainty that it provides little discriminatory power.

An alternative kinematic estimator is the
\textit{expansion age}, which one determines using a method
analogous to
the measurement of Hubble flow: one plots the positions 
of stars against their velocities in the same direction. 
If the stars are expanding,
their positions and velocities
will be correlated, and the slope of the correlation is just 
the inverse of the time since expansion began. 
\citet{torres_search_2006} 
apply this method to the $X$ positions and velocities 
of $\beta$PMG stars
to obtain an age of $\sim18\Myr$.\footnote{
Here and throughout we
adopt a standard $XYZ$ right-handed Cartesian coordinate system where
the Sun's position projected onto the Galactic plane
 lies at the origin in position, the Local Standard of Rest lies at the origin in velocity, and, at the origin, the positive $X$ direction is toward the Galactic centre, the positive $Y$ direction lies in the plane aligned with the direction of Galactic rotation, and the positive $Z$ direction is orthogonal to the Galactic plane. We use $U=\dot{X}$, $V=\dot{Y}$, and $W=\dot{Z}$ to denote velocities in this coordinate system, with  $U=V=W=0$ corresponding to the local standard of rest. As the coordinate system evolves through time it corotates as the origin travels along its circular orbit around the Galaxy, maintaining the axes directions as defined above.}
While this is less 
than $2 \sigma$ from the combined LDB and isochronal 
age of $23\pm3\Myr$, 
\cite{mamajek_age_2014} 
point out that the expansion slope is not consistent across 
dimensions. Indeed performing the same analysis in 
the $Z$ direction yields a negative slope, implying
contraction rather than expansion.

The problems in current kinematic techniques likely have two distinct causes. First, the methods are not robust when applied to moving groups whose origin sites have complex structures in space or time. For example, \citet{wright_kinematics_2018} investigate the expansion 
rate of the Scorpius-Centaurus OB Association (Sco-Cen 
hereafter) by assuming it could be decomposed into three distinct subgroups (despite evidence that the true structure is significantly more complex, e.g. \citealt{rizzuto_new_2015}), but find that the kinematics are more consistent with contraction than expansion, so that any expansion or traceback age one might derive is meaningless. Even in cases where stars are expanding, both traceback and expansion methods are likely to yield misleading results if there is a non-negligible spread in the spatial distribution or age of formation sites.

A second problem is membership determination. In order to 
apply a kinematic ageing technique to a moving group, one 
must start with a list of its members, constructed either 
by hand or using an automated tool such as LACEwING 
\citep{riedel_lacewing:_2017} or BANYAN 
\citep{gagne_banyan._2018} that assigns membership 
probabilities based on fits in 3D position or 6D phase 
space. Using hand-selected membership lists often produces 
results that depend significantly on which stars are 
included. However, with the automated methods the process is 
somewhat circular: the centre and dispersion of a 
purported moving group depends on which stars are included 
as probable members, but which stars are included in turn 
depends on the adopted centre and dispersion of the group. 
When one attempts a kinematic traceback using member lists 
determined in this fashion, the errors compound to the 
point where the method is not viable. 
\citet{riedel_lacewing:_2017} find that they can not 
determine kinematic ages for any known association, or 
even for a synthetic association described by a single age 
and a Gaussian distribution in space.

In this paper we introduce a new method called
\texttt{Chronostar} that addresses many of the problems 
discussed above. Compared to existing methods, 
\texttt{Chronostar} has several advantages: (1) it 
simultaneously and self-consistently solves the problems 
of membership determination and kinematic ageing; (2) it 
does not assume or require that moving groups have a 
single, simple origin in space and time, and thus 
allows for a more realistic representation of the complex 
structure of star-forming regions; (3) it uses forward 
modelling rather than traceback, thereby eliminating the 
need for complex and uncertain propagation of 
observational errors. Our layout for the remainder of this 
paper is as follows. We present the formal derivation of 
our method in \autoref{sec:methods}, and in 
\autoref{sec:synth_test} we test it on a variety of 
synthetic datasets, demonstrating that it is both robust 
and accurate. In \autoref{sec:bpmg} we present a simple 
application to the $\beta$ Pictoris Moving Group, showing 
that, for the first time, we are able to recover a 
kinematic age with tight error bars that is consistent 
with ages derived from other methods. We discuss
\texttt{Chronostar}'s performance in comparison with
other methods in \autoref{sec:discussion}.
Finally, we 
summarise and discuss future prospects for our method in 
\autoref{sec:conclusion}.
The code for \texttt{Chronostar} can be found
at https://github.com/mikeireland/chronostar.

\vspace{1cm}
\section{Methods}
\label{sec:methods}
\subsection{Setup}
\label{sec:methods-set-up}

\begin{figure}
\centering
\includegraphics[width=\columnwidth]{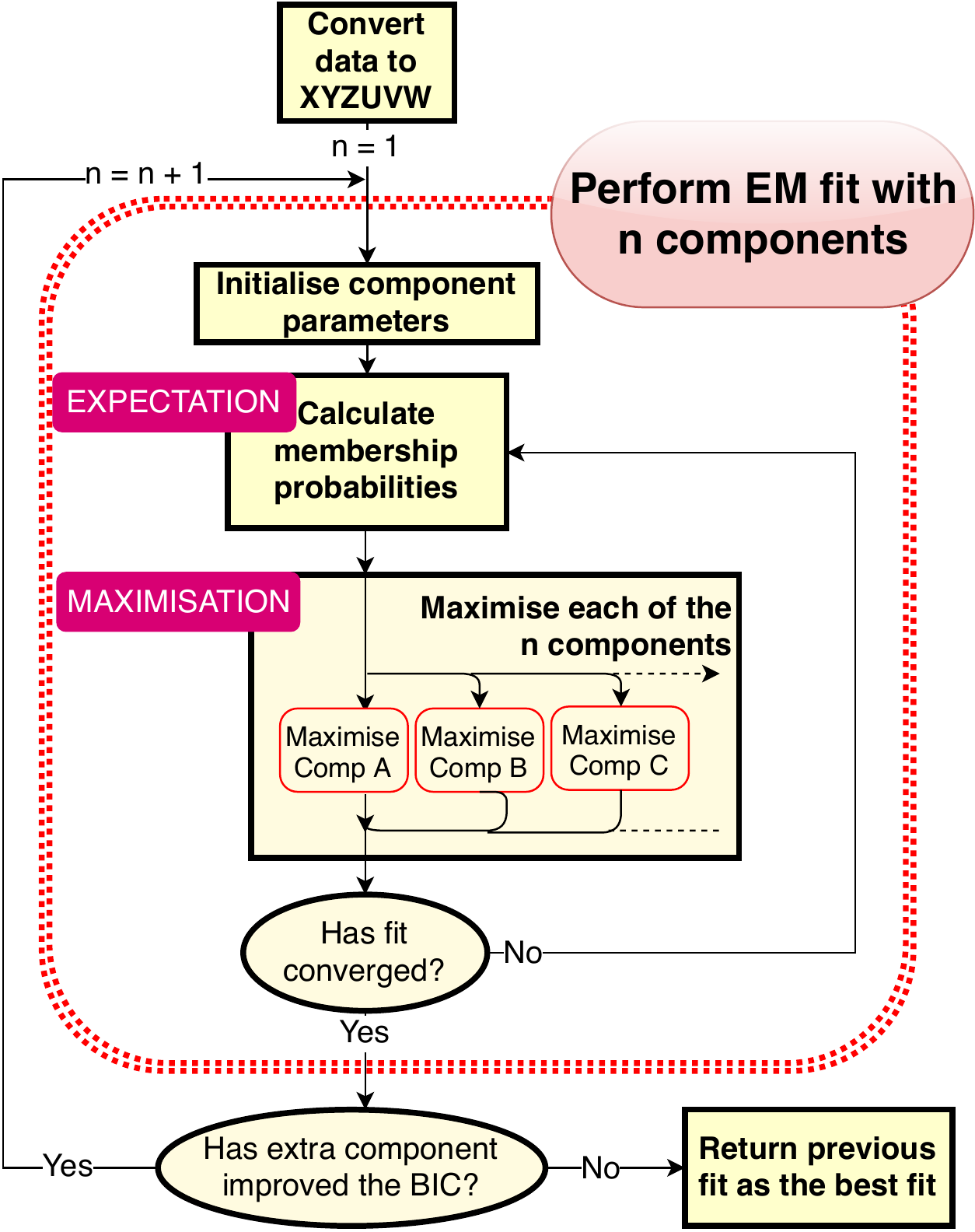}
\caption{Layout of the overall fitting algorithm to a set of stellar
6D astrometric data. We first convert the data into Cartesian
coordinates $XYZUVW$, centred on the local standard of rest.
We then perform an Expectation Maximisation (EM) fit (detailed
in \autoref{sec:em}), initially with the number of components
(described in \autoref{sec:single_component}) 
$n$ set to $1$, then subsequently incremented until the inclusion
of an extra component worsens the fit, as determined by the Bayesian
information criterion (BIC).
As part of the maximisation step, we maximise each component's set
of parameters using Markov Chain Monte Carlo sampling, by first
constructing a Gaussian distribution in 6D phase-space as defined
by the parameters, then projecting it forward through time by the 
modelled age, before comparing it to its assigned members
(see \autoref{fig:schematic}).
}
\label{fig:algorithm}
\end{figure}

\begin{table*}
\centering
\begin{tabular}{l|l|p{13cm}}
\hline
Symbol & Units & Meaning \\
\hline
$X,Y,Z$ & $\un{pc}{}$ & Positional cartesian dimensions, centred on 
the Sun's position projected onto the plane of the Galaxy,
positive towards Galactic centre, circular rotation and Galactic North respectively. \\
$U,V,W$ & $\kms$ & Cartesian velocity dimensions, centred on the local standard of rest,
with same orientation as $X,Y$ and $Z$ respectively.\\
$\vt$ & - & 6D phase-space position in $XYZUVW$. \\
$\Norm{\vt}{\vm}{\bS}$ & - &
Evaluation of the 6D Gaussian with mean $\vm$ and covariance
$\bS$ at the phase-space point $\vt$.\\
$\vm_0$ & - & Modelled centroid of the 6D Gaussian distribution in 
$XYZUVW$ representing the initial kinematic distribution of a component of an association. \\
$t$ & $\,$Myr & Modelled age of a component of an association. \\
$\sigma_{xyz}$ & $\,$pc & Modelled standard deviation in $X$, $Y$ and $Z$
of initial distribution. \\
$\sigma_{uvw}$ & $\kms$ & Modelled standard deviation in $U$, $V$ and $W$
of initial distribution. \\
$\bS_0$ & - & Modelled covariance matrix of a 6D Gaussian distribution in
 $XYZUVW$ representing the initial kinematic
 distribution of a component of an association, constructed from $\sigma_{xyz}$ 
and $\sigma_{uvw}$.\\
$C$ & - &
	A component modelled as a 6D Gaussian in phase-space
    defined by 9 parameters: initial phase-space
    centroid ($x_0,y_0,z_0,u_0,v_0,w_0$), initial standard deviations
    in position and velocity space ($\sigma_{xyz},\sigma_{uvw}$) and time
    since becoming gravitationally unbound, $t$.\\
$p(d|M)$ & - & The likelihood (unscaled probability) of seeing 
data point $d$ assuming that $d$ was drawn from the modelled
distribution $M$.\\
$\vf(\vt, t)$ & - & Abstracted function from \texttt{galpy} that 
numerically integrates the orbit of $\vt$
through the Galactic potential as a function of time $t$.\\
$\Omega_{i,k}$ & - & The overlap integral of the $i$th star with the
$k$th component. We calculate this by integrating over the convolution
of the two associated 6D Gaussians.\\
$\mathbf{Z}$ & - & Two-dimensional array of membership probabilities 
with a row for each star and a column for each component.\\
$w_k$ & - & Expected fraction of stars belonging to component $k$. We calculate
this by summing the $k$th column of $\bZ$, and normalising by the total number of stars. \\
\hline
\end{tabular}
\caption{Variables and parameters used in \autoref{sec:methods}. }
\label{tab:symbols}
\end{table*}

Our ultimate goal is to find
the most likely kinematic description of an association's origin,
such that evolving it through time by its modelled age to its
\textit{current-day} distribution, best explains the observed 
kinematic distribution
of the association's members.
In this section we detail our Bayesian approach to finding
the best kinematic
description of a stellar association
by modelling its origins as the sum of
Gaussians in 6D Cartesian
phase-space with independent ages,
means, and covariance matrices. We refer to each
Gaussian as a \textit{component},\footnote{A component
is a collection of stars with
similar 6D phase-space properties and similar age.
A simple association may only require a single
component, whereas an association with complex 
substructure like the Scorpius-Centaurus OB Association 
might be better described with multiple components.}
and for the simplified models presented here, each
component has the same standard deviation in each position
dimension, and the same standard deviation in each velocity
dimension.
We define the origin of an association as the 
point in time at which stars become gravitationally 
unbound and begin moving ballistically through the Galaxy.
We approximate the initial positions and velocities to be 
uncorrelated with one another, but as the association evolves 
in time those quantities will become correlated -- consequently 
the covariance matrix, and in particular the terms within 
it describing position-velocity covariance, 
are functions of time.
We fit these components to a set of observed stars
by maximising the overlap between the observed stellar 
position-velocity information, including the full error 
distribution and its covariances, and the Gaussian 
that describes the current-day structure of a component 
in phase-space. We include an assessment of membership
probability as part of this analysis.
We decide how many components to use to fit a given set
of stars by comparing the likelihood of the best fit in
each case using the Bayesian Information Criterion
(BIC). The BIC is a metric that balances the likelihood against a term that takes into account the number of parameters used to build the model. This term penalises the BIC (\citealt{schwarz_estimating_1978}) as more parameters are included, which lowers the chance of overfitting the data (see \autoref{eq:bic} and surrounding text for details).
We provide \autoref{tab:symbols} as a quick reference for
the variables and parameters introduced throughout this section.

We begin this section with a top-down description of the algorithm
(\autoref{fig:algorithm}).
We must first decide how many components to use in our fit.
A priori we do not know how many components are required to 
describe an association and so we run the following algorithm
iteratively, incrementing the number of components each time,
halting when the extra component yields a worse BIC value.
For a given component count we use the Expectation Maximisation (EM)
algorithm (e.g. \citealt{mclachlan_finite_2004})
to simultaneously find the best parameters of each of the
model components as well as the relative membership probabilities of 
the stars to each component.
After initialisation of the model parameters, EM iterates through the
Expectation step (E-step) and the Maximisation step (M-step) until
convergence is reached. The E-step consists of calculating membership 
probabilities to not only each of the components but also to the 
background field distribution. The M-step utilises the membership
probabilities to find the best parameters for each component through
the maximisation of an appropriate likelihood function. 

We now describe the model bottom up. We begin with the parametrisation
of a single component's origin as a spherical Gaussian 
(\autoref{sec:single_component}) and how we evolve this initial
distribution to its current-day distribution. Next we summarise 
the Bayesian approach to computing the likelihood function 
(\autoref{sec:fitting_approach})
for a single component. Finally we 
incorporate multiple Gaussian components using the EM algorithm,
as well as the background distribution, which is critical for 
accounting for interlopers.
\footnote{ The likelihood function provides a metric on 
how well a  given set of parameter values explains the data.}

\begin{figure}
\centering
\includegraphics[width=\columnwidth]{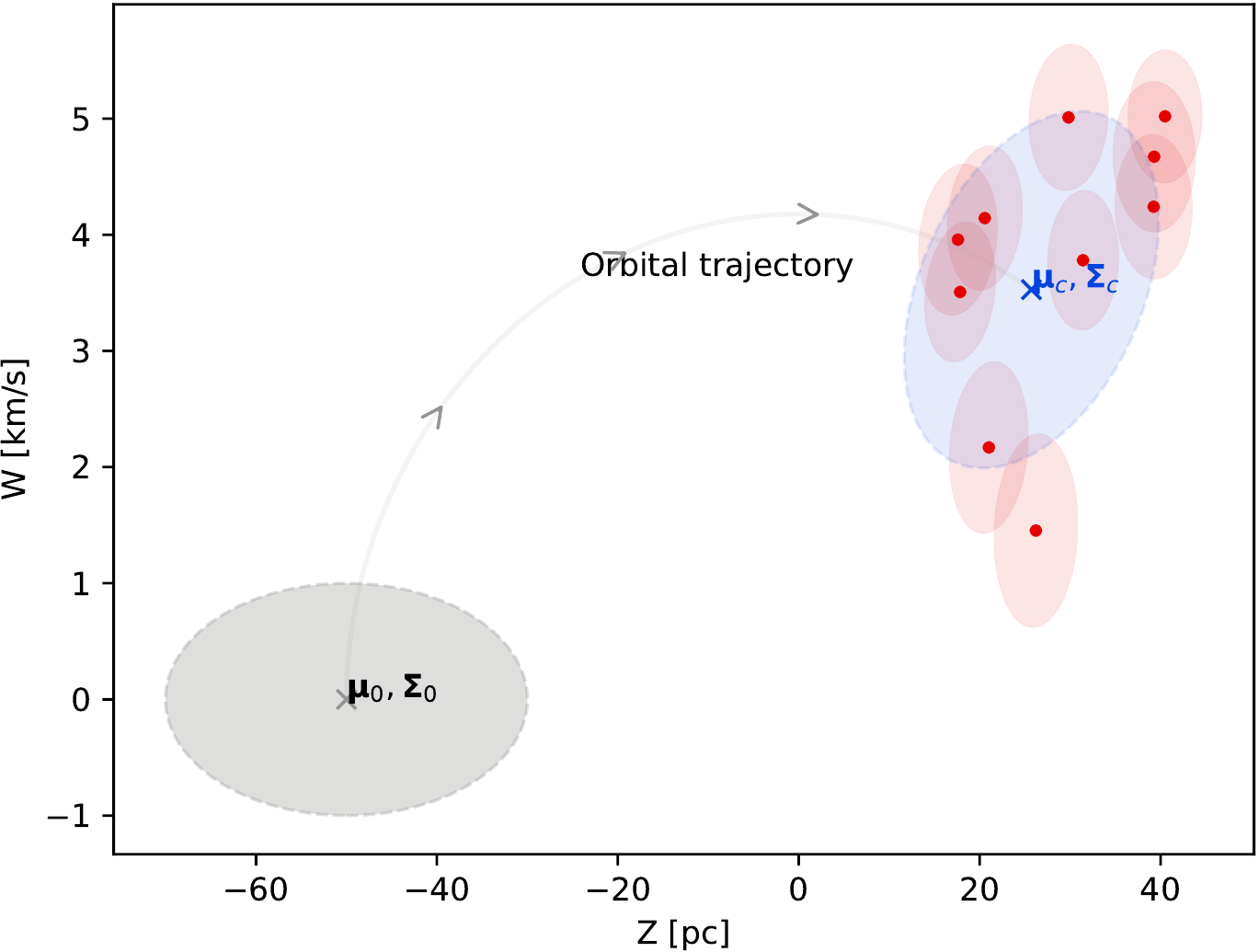}
\caption{A basic schematic detailing how a single
Gaussian component is fitted to stellar data.
The red points represent stellar data at current-day 
(e.g., as measured by \textit{Gaia} DR2)
with $2\sigma$
uncertainties denoted by ellipses. 
The grey ellipse and cross depict the modelled initial 
distribution, a 6D Gaussian with mean $\vm_0$ and 
covariance matrix $\bS_0$.
The fitting process projects the initial distribution forward
by the component's modelled age which generates
the current-day distribution with mean $\vm_c$ and 
covariance matrix $\bS_c$, depicted by 
the blue ellipse and cross respectively. The
orbital trajectory is denoted by the grey arc, with arrows
denoting travel forward through time.
For simplicity, we only show the $Z-W$ slice through the 6D
phase-space but \texttt{Chronostar} makes use of the
full 6D phase-space information.
}
\label{fig:schematic}
\end{figure}

The basic data on which our method will operate are a set 
of stars taken from \textit{Gaia} DR2  
\citep{gaia_collaboration_gaia_2018}.
For the purposes of this paper we focus on stars in
and around known associations, but in future work
the same algorithm can be applied
to search for new associations
and moving groups. 
We transform the 6D astrometry of each star
into 6D phase-space data $XYZUVW$, which describes
the position and velocity of the star in Galactic coordinates
as described in \autoref{tab:symbols}.
In addition to the 
central values for each star, we have an associated 
set of measurement errors encapsulated by a covariance matrix.
We use the transforms from \cite{johnson_calculating_1987}
to create a Jacobian from observed astrometry space to Cartesian
$XYZUVW$ space then use this to transform the
covariance matrices (similar to the process
detailed in \aref{ap:project}).
For the spatial coordinate origin we choose a point that coincides with the projection of the Sun onto the Galactic plane (the Sun being 25 pc above it) and whose velocity coordinate origin is given by the local standard of rest (LSR) as given by \citep{schonrich_local_2010}. For convenience we label this as
$\vt_{LSR} = [0,0,-25,-11.1,12.24,-7.25]$
with units as given in
\autoref{tab:symbols}, denoting our coordinate system origin with respect to the Sun. We apply this offset to the data to translate the initially heliocentric data to our chosen coordinate system. 

\subsection{Modelling a Single Component}
\label{sec:single_component}
As stated earlier, we use a spherical 6D Gaussian distribution
to model the origin of a component.
We define the kinematic origin of a collection of stars as
the approximation of some precise time and place when the stars
become gravitationally
unbound. A bound set of stars forms an ellipsoid in both position 
space and velocity space, with no correlation
between the three pairs of position and velocity dimensions ($X-U$, $Y-V$ and $Z-W$).
We refer to these three planes as \textit{mixed-phase planes} henceforth. We further
simplify matters by approximating the ellipsoid as spherical,
thereby removing all correlations between any dimensions.
We explore the validity of these assumptions in the discussion.

We parametrise the origin of a component
as a Gaussian in 6D phase-space $\vt=[x,y,z,u,v,w]$ 
with mean $\vm_0$ representing the vector of expectation values 
in each dimension:
\begin{equation}
\vm_0 = [x_0,y_0,z_0,u_0,v_0,w_0].
\end{equation}
To satisfy the criteria stated above we parametrise the covariance matrix
$\bS_0$ as:
\begin{equation}
\bS_0 = 
	\begin{bmatrix}
	\sigma_{xyz}^2 & 0 & 0 & 0 & 0 & 0 \\
	0 & \sigma_{xyz}^2 & 0 & 0 & 0 & 0 \\
	0 & 0 & \sigma_{xyz}^2 & 0 & 0 & 0 \\
	0 & 0 & 0 & \sigma_{uvw}^2 & 0 & 0 \\
	0 & 0 & 0 & 0 & \sigma_{uvw}^2 & 0 \\
	0 & 0 & 0 & 0 & 0 & \sigma_{uvw}^2 \\
	\end{bmatrix}.
\label{eq:cov_matrix}
\end{equation}
Note, that since we restrict our model to be separately
spherical in both position and velocity space
we can denote the initial standard deviation in each
position axis
(i.e. the radius of the association) as $\sigma_{xyz}$
and the initial velocity dispersion in each velocity
axis as $\sigma_{uvw}$.
Hence we can express the probability density associated with
each component as a
Gaussian distribution over $\vt$:
\begin{equation}
\label{eq:multi_gauss}
\mathcal{N}(\vt; \vm_0, \bS_0)
 = \frac{\exp\big[-\frac{1}{2}(\vt-\vm_0)^T\bS_0^{-1}(\vt-\vm_0)\big]}
 		{\sqrt{(2\pi)^6 |\bS_0|}}.
\end{equation}

In order to relate the distribution of a component to
observed stellar data we require the phase-space values of
both the  component model and the stellar data
 to be evaluated at the same
time. We 
transform the distribution of a component
from its origin (parametrised by $\vm_0$ and $\bS_0$)
forward through the Galactic potential by its modelled age
$t$, to its current-day distribution, another
Gaussian described by $\vm_c$ and $\bS_c$,
i.e. $\mathcal{N}(\vt;\vm_c,\bS_c)$ (see \autoref{fig:schematic}).
Using \texttt{galpy} \citep{bovy_galpy:_2015} to calculate orbits,
we transform the shape of the Gaussian by considering the orbital
projection of the distribution as a transformation between coordinate
frames. 
We use \texttt{galpy}'s model \texttt{MWPotential2014} as our model for the Galactic potential, but we show in \aref{ap:dependence} that choosing other plausible potentials does not lead to large differences in the results.
We can thus calculate the current-day distribution by
performing a first-order Taylor expansion about $\vm_0$, and 
generating the current-day covariance matrix $\bS_c$ under the approximation that this
coordinate transformation is linear 
 (details in Appendix~\ref{ap:project}).

\subsection{Fitting approach}
\label{sec:fitting_approach}
Now that we have the means to get the current-day distribution of a
component from its modelled origin point, we can use a Bayesian
approach to generate a probability distribution of the model's
parameter space, which will allow us to identify each parameter's
most likely value and associated uncertainty. 
As is standard with a Bayesian approach, we write the posterior
probability distribution of the model parameters ($C = \{\vm_0, \bS_0, t\}$)
given the data as the product of the prior probabilities with
the likelihood function:
\begin{equation}
p(C|D) \propto p_\text{prior}(C) p(D|C).
\end{equation}
The prior ($p_\text{prior}$) represents our initial guess at the
parameters in the absence of data, for example a restriction that
the initial spread, dispersion and approximate mass of the system
be super-virial (see \autoref{sec:priors} for details). The likelihood
function $p(D|C)$ is simply the probability density of the data
given the model.

In our context the data $D$ are composed of a set
of $N$ stars $\{s_1, s_2, \dots, s_n\}$ that are candidate members
of a component $C$, each with full 6D kinematic information.
From the method described in \autoref{sec:single_component}
we produce a current-day distribution ($\vm_c$, $\bS_c$) from the model
component parameters. We interpret the current-day Gaussian
as the probability density of finding a member
of $C$ at phase-space position $\vt$. 
If measurements were infinitely precise
this would be $\Norm{\vt}{\vm_c}{\bS_c}$
and the likelihood function for
a set of $N$ stars drawn independently would simply be 
$p(D|C) \propto \prod_i^N\Norm{\vt_i}{\vm_c}{\bS_c}$.
However, measurements of
$\vt_i$ have finite errors, which we take to be Gaussian, described by the probability distribution 
$\Norm{\vt}{\vm_i}{\bS_i}$, 
where $\vm_i$ is the central estimate and $\bS_i$ is 
the  
covariance of measurement errors. The likelihood product therefore becomes the product of convolutions of the Gaussian for component $C$ with the Gaussians describing the error ellipse for each star:
\begin{equation}
p(D|C) \propto \prod_{i=1}^N \int \Norm{\vt}{\vm_i}{\bS_i} \Norm{\vt}{\vm_c}{\bS_c} \, d\vt \equiv \prod_{i=1}^N \Omega_{i,c}
\label{eq:tf-likefun}
\end{equation}
where $\vm_i$ and $\bS_i$ are the central estimate and
covariance matrix for the $i$th star respectively, and
\begin{eqnarray}
\label{eq:Omega}
\Omega_{i,c} & = & 
\frac{\exp\left[-\frac{1}{2}\left(\vm_i-\vm_c\right)^T \bS_{i,c}^{-1}\left(\vm_i-\vm_c\right)\right]}{\sqrt{\left(2\pi\right)^6 \left|\bS_{ic}\right|}}
 \\
\bS_{i,c} & = & \bS_i + \bS_c.
\end{eqnarray}
\autoref{eq:Omega} is the standard result for the convolution of $N$-dimensional Gaussians; note that, in the limit of no errors ($\bS_i \rightarrow \mathbf{0}$), the result trivially reduces to $\Omega_{i,c} = \Norm{\vm_i}{\vm_c}{\bS_c}$ as described above. For convenience in what follows, we shall refer to $\Omega_{i,c}$ as the overlap integral of star $i$ with component $C$.

\subsection{Fitting many components with the Expectation Maximisation algorithm}
\label{sec:many}
\label{sec:em}
We now describe how to extend our formalisation so as to incorporate
multiple Gaussian components, i.e. a Gaussian Mixture Model.
Our model $M$ is a linear combination of 
$K$ components $\{C_1, C_2, \ldots, C_k\}$, and has the probability distribution function (PDF):
\begin{equation}
p(\vt|M) = \sum \limits_{k=1}^K w_k \Norm{\vt}{\vm_k}{\bS_k},
\end{equation}
where $w_k$ is the weighting of each component such that
$\sum_k w_k = 1$.
Intuitively $w_k$ is the expected
fraction of stellar members belonging to component $k$.
We calculate $w_k$ by summing the $k$th
column of $\bZ$ (defined below) and normalising by the total
number of stars.

To simplify the maximisation of the likelihood function, the common
approach is the Expectation Maximisation (EM) algorithm
\citep{mclachlan_finite_2004}. There are many derivations of this algorithm,
but for convenience we provide a brief summary.
The central problem in mixture models is how to assign particular stars to particular components. EM addresses this by introducing
a so-called \textit{hidden} variable $\bZ$ that tracks each star's membership
probabilities. $\bZ$ is a matrix of $N$
rows (for each star) and $K$ columns (for each component).
Each entry is a decimal number
between $0$ and $1$ such that each row sums exactly to $1$.
In this way the $ik$th 
element of $\bZ$ is the probability that the $i$th star
is a member of the $k$th component.

The expectation step (E-step) and maximisation step (M-step) have a
circular dependency: one cannot know the membership probabilities
without a fit to the components, and one cannot fit the components
without knowing which stars are members.
This is solved by, after a carefully chosen initialisation
(described below),
the algorithm alternating between the E-step (evaluating $\bZ$) and
the M-step (maximising each component's likelihood function) 
until convergence is achieved.
We can initialise this method by either using membership
probabilities from the literature to guess our $\bZ$, or by using
fits to the distribution from the literature to guess the model
parameters for the origin. We defer a more detailed discussion of how we initialise our fits to \autoref{ssec:adding_components}.

The E-step is the calculation of $\bZ$, for a fixed set of components. The  relative probability that star $i$ is a member of component $k$ with properties $C_k$ is 
 given by its overlap integral with that component scaled by 
the component weight $w_k$, so the total probability that it is 
a member of component $k$ is
\begin{equation}
Z_{ik} = \frac{w_k \Omega_{i,k}}{\sum_{j=1}^K w_j \Omega_{i,j}},
\end{equation}
where we use $\Omega_{i,k}$ to denote the overlap integral 
(\autoref{eq:Omega}) evaluated using $\vm_c = \vm_{k}$ and 
$\bS_c = \bS_k$, i.e., using the central location and 
covariance matrix for component $k$.

The M-step maximises the  likelihood function for fixed 
membership probabilities. By the introduction of 
$\bZ$, the likelihood function becomes separable,
allowing us to maximise each component's contribution in isolation.

The likelihood function for each component is the same as the 
likelihood function evaluated for a single component 
(\autoref{eq:tf-likefun}), modified so that each star is 
weighted by the probability that it is a member:
\begin{equation}
p(D|C_k,\bZ) \propto \prod_{i=1}^N  (w_k \Omega_{i,k})^{Z_{ik}}
\label{eq:one_of_many_lnlike}
\end{equation}
In the M-step, we use Markov Chain Monte Carlo (MCMC) (implemented
by \citealt{foreman-mackey_emcee:_2013} as \texttt{emcee})
to find the maximum likelihood values of $\vm_k$ and 
$\bS_k$ given the likelihood function $p(D|C_k, \bZ)$ 
and the priors we place on $C_k$ (see below). We reset 
$\vm_k$ and $\bS_k$ to these maximum likelihood values 
and then return to the E-step for another iteration. The
algorithm continues in this manner until converged, which we take
to be when the previous iteration's best fit parameters all fall
within the central 70th percentile of the new fit's
respective posterior distribution.

\subsection{Priors}
\label{sec:priors}
Our Bayesian treatment of data requires a prior on the parameters 
$t_k$, $\vm_{k,0}$ and $\bS_{k,0}$ that describe each component.
We first discuss our \textit{non-informative} priors. 
We have a uniform prior on $t_k$ and each element in $\vm_{k,0}$.
The covariance matrix $\bS_{k,0}$ is parametrised by only two values, $\sigma_{xyz}$ and
$\sigma_{uvw}$, both of which are standard deviations.
Since standard deviations are by definition restricted to be
positive, 
the natural non-informative prior is uniform in $\ln\sigma$ rather than in $\sigma$ itself, which corresponds to $p_\text{prior}(\sigma) \propto 1/\sigma$. 
Therefore, our prior on  $\bS_{k,0}$ is

\begin{equation}
p_\text{prior}(\bS) \propto p_\text{prior}(\sigma_{xyz}) p_\text{prior}(\sigma_{uvw})
\propto \frac{1}{\sigma_{xyz} \sigma_{uvw}}.
\label{eq:prior}
\end{equation}

We also add an informative prior regarding the dynamical state of purported origin sites. In testing we found that there is a mild degeneracy shared by the initial spatial volume of an association and its age, whereby some fits would collapse
to an extremely small $\sigma_{xyz}$. This is unphysical, since such a tightly packed association would have been gravitationally bound and thus would not have dispersed in the first place. To counter this, we introduce
a prior on the  virial ratio $\alpha$ of our components, which we approximate as
\begin{equation}
    \alpha = \frac{2\sigma_{uvw}^2 \sigma_{xyz}}{G M},
\end{equation}
where $M$ is the mass of the proposed component and $G$ is
the gravitational constant.
The mass of the association is not precisely known, since 
often the masses of individual stars are constrained 
only poorly, and our lists of candidate members are 
magnitude-limited and thus likely omit a significant 
number of low-mass stars. As a very rough estimate we 
adopt $M = nM_\odot$ for the purpose of computing 
$\alpha$, where $n$ is the number of stars in a 
proposed component. This amounts to assuming
solar mass stars.
We apply a prior with a Gaussian distribution
on the natural logarithm
of $\alpha$ with a mean of $2.1$ and a standard
deviation of $\sigma_\alpha =$ $1.0$. We select these values
such that the mode of the corresponding log-normal
distribution occurs at $\alpha=3$.
We find that smaller values of $\sigma$$_\alpha$ smoothly increase
the fitted age when applied to real data
as the fit prioritises more compact origins regardless of the data,
whilst the age fit converges for
$\sigma_\alpha \geq 1$.

\subsection{Characterising field stars}
\label{sec:characterise}

For any proposed member of an association there
is always a chance it is not truly a member, but rather
an interloper that happens to share similar kinematic 
properties. We label these stars as \textit{field} stars,
and call the PDF of all field stars the \textit{background}
distribution.
We use the background distribution to consider the probability
that a star is a member of the background, and thus 
properly quantify the star's membership
probability to our association fits.

We determine our background distribution, which is held
fixed, as follows. We select all \textit{Gaia} DR2 stars with radial velocities and parallax
errors better than 20\% ($n$=6,376,803 stars), and then transform the data into 
Galactic cartesian coordinates ($XYZUVW$) as described in \autoref{sec:methods-set-up} .
We estimate the background PDF shape in these coordinates using 
a Gaussian Kernel Density Estimator (KDE) to approximate
a continuous PDF. The KDE we use\footnote{\texttt{scipy.stats.gaussian\_kde} \citep{jones_scipy:_2001}} represents the PDF as a sum of Gaussians centred on each of the $n$ input data points; these Gaussians have a covariance matrix equal to the covariance matrix of the input data set, scaled down by the square of a dimensionless factor called the \textit{bandwidth}. To be precise, we estimate the background stellar density at a point $\vt$ in parameter space as
\begin{equation}
    p(\vt) = \sum_i \mathcal{N}(\vt; \vm_i, h^2 \bS_G),
\end{equation}
using notation from \autoref{eq:multi_gauss}, where the sum runs over the $n$ stars in the \textit{Gaia} DR2 catalog with acceptably small errors, $\bS_G$ is the covariance matrix of this catalog, $\vm_i$ is the $XYZUVW$ position of star $i$, and $h$ is the bandwidth. We determine $h$
using Scott's Rule \citep{scott_multivariate_1992}:
$$h = n^{-1/(d+4)} $$
where $d=6$ is the 
number of dimensions, resulting in a bandwidth of $\approx0.2$.

We treat the background as simply another component 
in our multi-component fits, i.e., in a fit with $K$ 
components, the parameter $\bZ$ is an
$N\times (K+1)$ 
matrix, with the final column $K+1$ giving the 
probability that a given star is a background star 
rather than a member. Our treatment of the background 
differs from that of other components only in that 
the background is static and does not have any 
parameters that can change, so the overlap integral 
between it and each star $\Omega_{i,K+1}$, is a 
constant that may be computed once at the beginning 
of our calculation and then stored for use when 
needed. We further note that the background is 
essentially constant over scales in $XYZ$ and $UVW$ 
comparable to the sizes of measurement uncertainties, 
and thus we can approximate the overlap integrals 
$\Omega_{i,K+1}$ as simply the value of the 
background evaluated at the central position and 
velocity estimates for each star.

\subsection{Initialisation and adding components}
\label{ssec:adding_components}

Each run of \texttt{Chronostar} begins with a single component, 
which is described by nine scalar quantities: the six 
components of the central phase-space position $\vm_0$, the 
initial spatial and velocity dispersions $\sigma_{xyz}$ and 
$\sigma_{uvw}$, and the age $t$. We initialise an MCMC search 
for the maximum likelihood in the nine-dimensional space these 
parameters describe by placing walkers randomly around a 
central starting guess, which is that the components of $\vm_0$ 
are equal to the mean of the stellar data to which we are 
fitting, $\sigma_{xyz} = 20$ pc, $\sigma_{uvw} = 7 \kms$, and 
$t = 3$\,Myr. We then run the MCMC algorithm to maximise the 
likelihood function as described in \autoref{sec:em}, stopping 
when we reach convergence. At this point we have predicted 
posterior probability distributions for all nine quantities.

To trial an alternative two-component fit, we must specify a 
starting guess for the parameters of each of the two 
components, which will serve as starting points about which to 
distribute initial MCMC walker positions. We choose these
starting guesses as follows. For each proposed component, we 
take our starting guesses for $\sigma_{xyz}$ and $\sigma_{uvw}$ 
to be equal to the maximum posterior probability value 
derived from the one-component fit. We set the starting 
time guesses for our two proposed components equal to the 16th 
and 84th percentile values of the posterior probability 
distribution for $t$ from the one-component fit. Finally, we 
set our starting guess for the phase-space position $\vm_0$ of 
each of the two trial components such that their 
\textit{current-day} positions match the \textit{current-day} 
centroid position of the maximum posterior probability value 
for the one-component fit. 
From these starting guesses, we run the EM algorithm as 
described in \autoref{sec:em}, stopping when we converge.
We  then compute the BIC for the one-component versus the 
two-component fit  using the following formula:
\begin{equation}
    \color{blue} \text{BIC} = \ln{(n)}k - 2\ln{\mathcal{L}},
    \label{eq:bic}
\end{equation}
where $n$ is the expected number of stars assigned to the components, $k$ is the number of parameters, and $\mathcal{L}$ is the evaluated likelihood.
If the two-component BIC is inferior to 
the one-component result, we stop and accept the one-component 
result. If not, we accept the two-component fit, and consider 
the possibility of adding a third component.

The procedure for going beyond two components is much the same 
as for going from one to two: we initialise a search by 
splitting an existing component into its 16th and 84th 
percentile ages, setting our central guesses for the initial 
phase-space positions and dispersions exactly as we did when 
going from one to two components. The primary complication is that 
there is no obvious means to identify which existing component 
should be split in order to yield the best fit. Thus 
\texttt{Chronostar} explores all possible splits, carries out 
EM to find the posterior probability for each possibility, then 
selects the best option between these and the previous fit to 
one fewer components based on the BIC. We stop adding 
components when the fit with $N$ components has a superior BIC 
compared to any of the possible fits using $N+1$ components.

We caution that this procedure does not necessarily guarantee 
the identification of the best possible fit, especially when 
the number of components is relatively large. As is usual with 
MCMC over high-dimensional spaces (a fit to $N$ components has 
$9N$ dimensions), there is in general no way to guarantee that 
the true global maximum likelihood has been found. However, 
\autoref{sec:synth-multi-comps}  demonstrates that even this 
basic approach successfully decomposes complex associations 
with multiple components.

The time taken to complete a fit is strongly dependent on the number of components required, but also depends on the number of stars and the age of the components.  Each EM iteration takes on average 500-1000 MCMC steps per component. The number of EM iterations needed for convergence can vary from about 30 iterations (in the case where there is a clear sub-component needing characterisation) to upwards of 150 iterations (where the introduced component has not identified a separate distinct over-density). Simple arrangements, such as the multi-component tests we present in the next section, require only $\sim 10$ hours for convergence running on a workstation or a single node of a cluster. A blind fit to $\sim 2000$ stars will take upwards of a week on the same hardware. In the case of our $\beta$PMG fit (see \autoref{sec:bpmg}), we ran the fit with 19 threads (1 per each of the 18 walkers plus one master thread)\footnote{Each component could in principle be fitted concurrently in the maximisation stage. However, we have thus far not implemented parallelism in this step, so each component is maximised sequentially.} and the computation took about a week to converge. Due to these limitations, if \texttt{Chronostar} were to fit to all of \textit{Gaia}, the data would need to be parceled into subsets of a few thousand stars at a time.

\section{Testing Chronostar with synthetic data}
\label{sec:synth_test}
We investigate the reliability and accuracy of our fitting
approach by testing it on 
an extensive suite of 1080 synthetic single-component 
associations, three scenarios with multiple components,
a set of stellar kinematics taken from a star formation
simulation (\citealt{federrath_inefficient_2015}; 
\citealt{federrath_converging_2017}),
and a two-component association within a uniform background.
We describe each of these tests in turn in the following sections.

\subsection{Single-component analysis}
\label{sec:synth_gen}

Our first test is the simplest, and uses as its mock 
data single components constructed from the same 
distributions we have assumed can be used to describe 
real associations. Thus we consider origin sites that
are spherical in both position and velocity space, and
are parametrised by 5 values: age $t$ (i.e., how many years have
passed since becoming unbound), $\sigma_{xyz}$, the 
standard deviation in each position dimension,
 $\sigma_{uvw}$, the velocity dispersion or standard deviation
in each velocity dimension, 
$n$, the number of stars drawn from the distribution,
and $\eta$, which characterises the uncertainty of the
observed current-day properties as described below.
The purpose of this test is to ensure that our code 
can recover input data whose properties match our 
assumptions, and to characterise the level of 
accuracy we can expect in this optimal case.

We use our chosen parameters to construct a synthetic data
set in several steps, which we explain in more detail below: 
(1) we use $t$, $\sigma_{xyz}$, and $\sigma_{uvw}$ to compute
the starting centroid position $\mu_0$ and covariance matrix $\bS_0$ for
our synthetic association; (2) we use $\vm_0$ and $\bS_0$ plus
the specified number of stars $n$ to create a set of synthetic initial
positions and velocities, which we then integrate forward by time $t$
to produce a set of synthetic current-day 
positions and velocities; (3) we add synthetic 
errors, whose sizes are parameterised  by $\eta$, to 
yield a set of ``observed'' stars on which we run 
the  \texttt{Chronostar} code.

Step (1) is to compute $\mu_0$ and $\bS_0$. For the latter we set
$\bS_{00} = \bS_{11} = \bS_{22} = \sigma_{xyz}^2$ and $\bS_{33} = \bS_{44} = \bS_{55}
= \sigma_{uvw}^2$, following the assumption that our initial conditions are
spherical distributions; all off-diagonal components of $\bS_0$ are zero.
For the former, we choose a
starting position so that the current-day centroid position of our synthetic
association matches that of
Lower Centaurus-Crux (LCC), $\vm_{\text{LCC}} = [50,-100,25,1.1,7.76,2.25]$ (see \autoref{tab:symbols} for units) by integrating an orbit backwards through time for the desired age $t$, beginning at the desired current-day centroid $\vm_c$.
In other words: $\vf(\vm_0, t) = \vm_{\text{LCC}}$, where
$\vf(\vm_0,t)$ is the function that maps an initial position $\vm_0$
to a final position $\vm$ after orbiting a time $t$ through the Galactic
potential.  This choice ensures the synthetic 
associations are all at the same heliocentric radius 
thus preserving consistency with measurement 
uncertainties that are distance dependent.

Step (2) is drawing $n$ stars from
a 6D Gaussian distribution with centroid $\mu_0$ and covariance matrix $\bS_0$.
We then integrate these stars forward through
the Galactic potential for a time $t$.
We convert the current-day position of each star from Cartesian Galactic coordinates to
astrometric coordinates (RA, DEC, $\mu_{\rm RA}$, $\mu_{\rm DEC}$,
parallax and radial velocity).

\begin{figure}
	\includegraphics[width=\columnwidth]{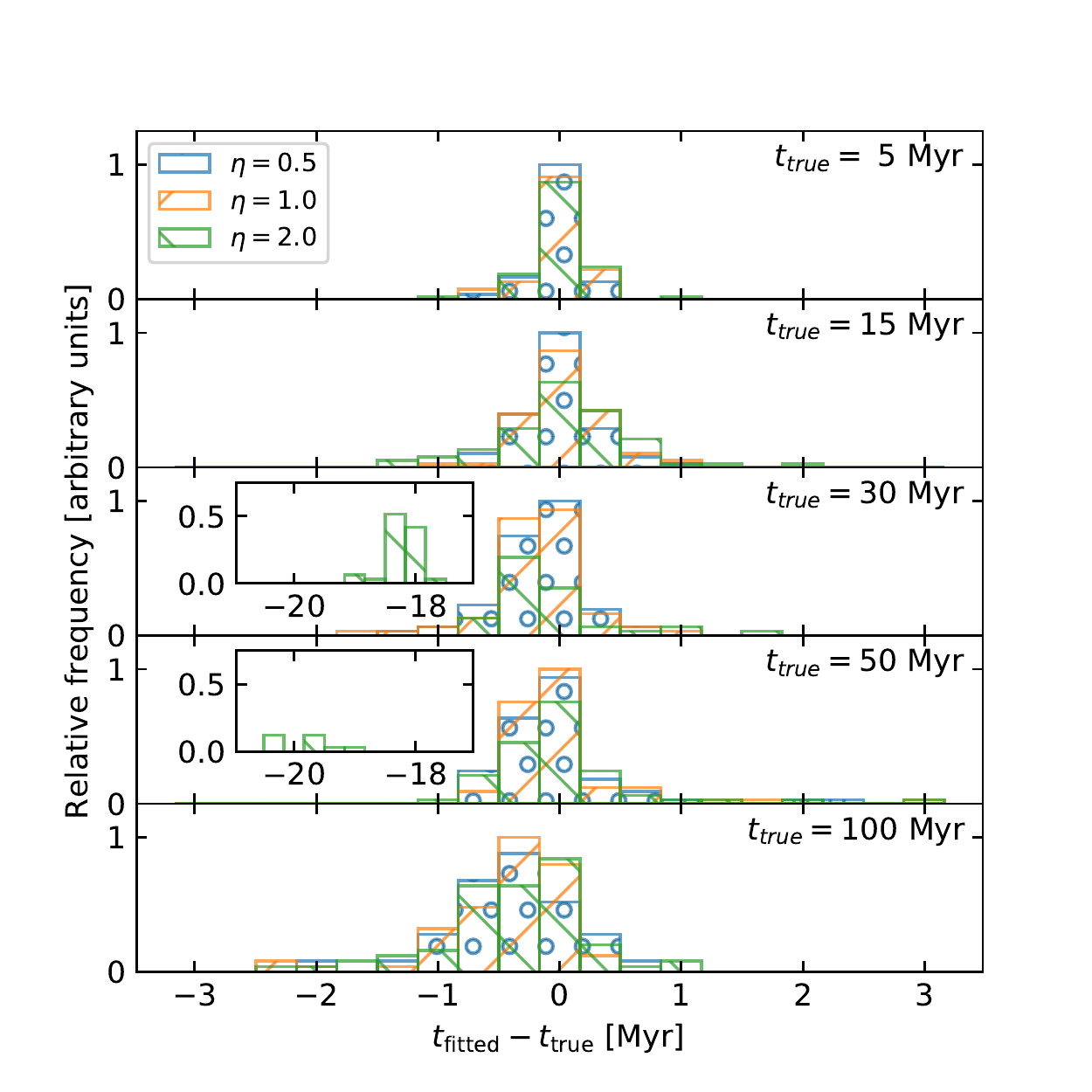}
	\caption{
    Histograms of residuals $t_\text{fitted} - t_\text{true}$ 
	resulting from application of
    \texttt{Chronostar} to 1080 synthetic associations, 
    grouped by true age
    (panels, increasing from top to bottom) and degree
    of measurement uncertainty (histograms in each panel).
    The catastrophic failures (featured in the insets) all cluster around 20\,Myr, which
    is equal to the quarter period of vertical oscillations
    through the Galactic plane.
    }
	\label{fig:raw-resids-step}
\end{figure}

\begin{figure}
	\includegraphics[width=\columnwidth]{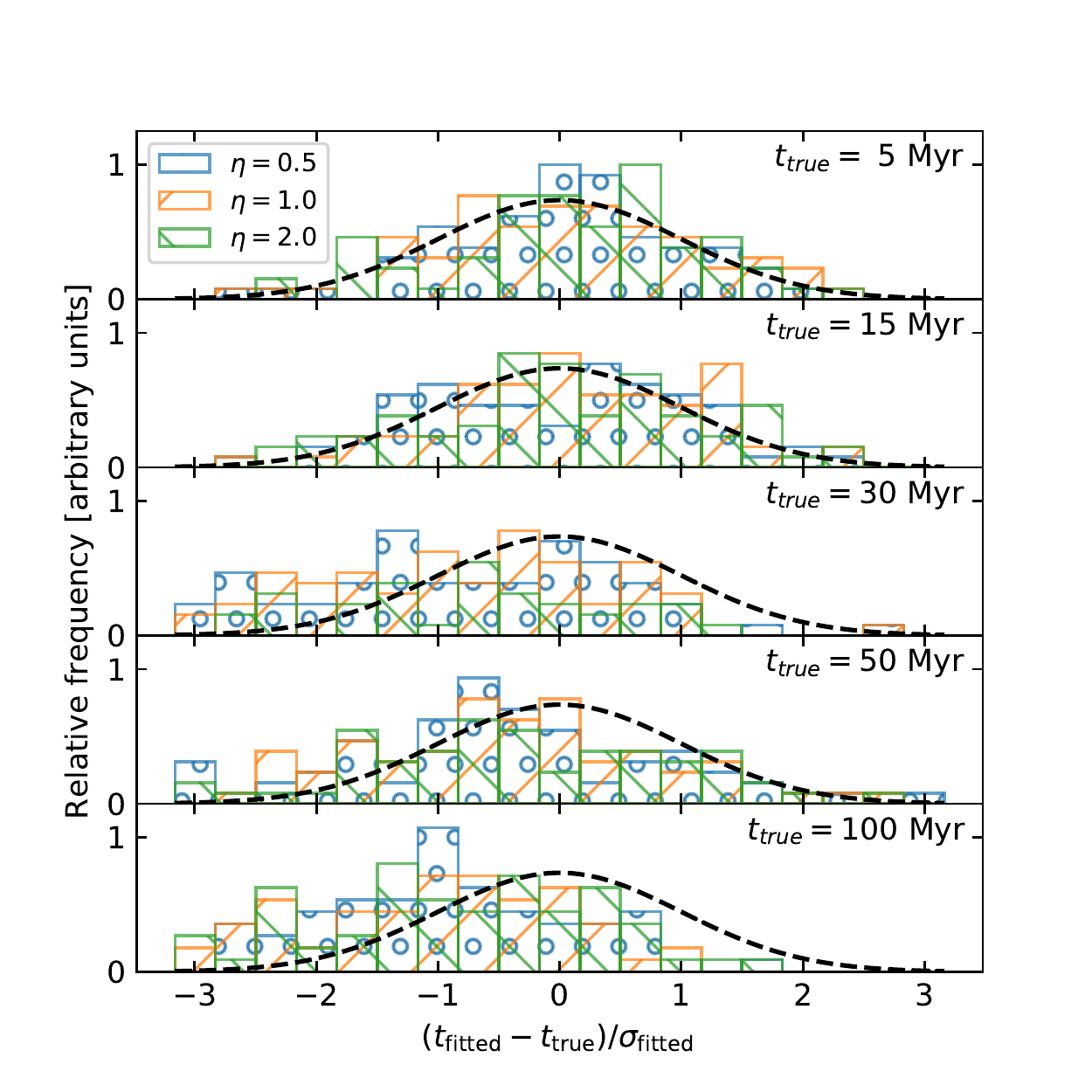}
	\caption{
    Same as \autoref{fig:raw-resids-step}, except that we show the
    distribution of normalised residuals 
    $(t_{\rm fitted}-t_{\rm true})/\sigma_{\rm fitted}$ rather than the
    distribution of raw residuals. For comparison, the black lines in
    each panel show Gaussian distributions with zero mean and unit variance,
   scaled such that the area under the curve matches the area under
    the $\eta=0.5$ histogram in each panel,
    and thus shows the distribution of normalised residuals we
    would expect if the true posterior PDF were Gaussian.
    We do not include the catastrophic failures in this figure.
    }
	\label{fig:normed-resids-step}
\end{figure}

Step (3) is to add synthetic errors.
We use the median uncertainties of \textit{Gaia} DR2 to inform our
artificial measurement uncertainties. Of all the \textit{Gaia} DR2
stars with radial velocities and parallax uncertainty better
than 20\%, the median uncertainties for parallax,
proper motion and radial velocity are $0.035$~mas,
$0.05$~mas~yr$^{-1}$ and 1~km~s$^{-1}$ respectively.
We set the uncertainties on our synthetic data by
multiplying these values by a dimensionless scale factor $\eta$,
so that the parallax uncertainty is $0.035\eta$ mas, the proper
motion uncertainty is $0.05\eta$ mas yr$^{-1}$, and the radial
velocity uncertainty is $1\eta$ km s$^{-1}$.
For each synthetic star, we add a random offset to its astrometric
coordinates chosen by drawing from a Gaussian distribution of the
specified size.

We generate 1080 synthetic associations by creating 
four realisations for each possible combination of the parameters:
\begin{align*}
\text{age},\, t&= [5, 15, 30, 50, 100] \un{Myr}{}, \\
\text{radius},\, \sigma_{xyz} &= [1, 2, 5] \un{pc}{}, \\
\text{velocity dispersion},\, \sigma_{uvw} &= [1, 2] \kms, \\
\text{star count},\, n &= [25, 50, 100],\\
\text{error scaling},\,\eta & = [0.5, 1, 2].
\end{align*}
We choose this range of parameters to be
broadly representative of known or claimed moving group origin sites. 
However, the maximum values of $\sigma_{uvw}$ and $t$ for which we 
test deserve special mention, because they are driven in part by the
limitations of our method. In deriving the likelihood function we 
have approximated the time evolution of the PDF of phase-space density 
with a linear transformation, and we show below that
this approximation breaks down for association ages $\gg 100$\,Myr,
or velocity dispersions $\gg 2$ km s$^{-1}$.

For each synthetic association we 
run \texttt{Chronostar} using a single component; the output
of the run is a set of MCMC walker positions in the space 
$(t, \vm_0, \sigma_{xyz}, \sigma_{uvw})$. From these positions, we
define $t_{\rm fitted}$ as the median $t$ coordinate of the walkers,
and the corresponding fit uncertainty $\sigma_{\rm fitted}$ as half
the difference between the 16th and 84th percentiles of $t$.

\autoref{fig:raw-resids-step} shows the distribution of
raw residuals, $t_{\rm fitted} - t_{\rm true}$ (where $t_{\rm true}$
is the true age used to generate the synthetic data set) that
we obtain from our experiment, and \autoref{fig:normed-resids-step} shows
the corresponding normalised residuals, 
$(t_{\rm fitted} - t_{\rm true})/\sigma_{\rm fitted}$; 
in both cases the data are grouped by values
of $t_{\rm true}$ and $\eta$. The raw residuals characterise the
absolute accuracy of the method, while the distribution of
normalised residuals characterise the accuracy of the error estimate
it returns.

We find that, except for a small number of
catastrophic failures that are easy to spot and
discussed below, \texttt{Chronostar} recovers
the correct ages to accuracies of $\sim 1$\,Myr almost independent
of $t_{\rm true}$ or $\eta$. There 
is a systematic bias toward younger
ages that increases with $t_{\rm true}$, reaching a maximum
net offset of $\sim0.3\Myr$ at 
$t_{\rm true} =100\Myr$.
We attribute this offset to the associations' minor yet potential departures
from the linear regime.
The normalised residuals show distributions
that are close to Gaussians with unit dispersion, indicating that the
error distribution for our method is close to Gaussian, and that the
returned $\sigma_{\rm fitted}$ is an accurate estimate of the true
uncertainty. Again we see a slight bias toward 
younger ages that worsens for older $t_\text{true}$.

It may seem surprising that the age fits have an uncertainty so robust to 
$t_{\rm true}$. However ultimately the uncertainty of the age fit
depends on how accurately \texttt{Chronostar} fits the correlation 
in the three mixed phase-space planes 
($X-U$, $Y-V$ and $Z-W$), as 
these are the signatures of expansion. The stars mostly remain in 
the linear regime so these correlations are linear regardless of the 
age, therefore the reliability of the age fit is almost independent of age.

\autoref{fig:raw-resids-step} shows that our method catastrophically
fails for a small number of cases where the true age is 30 or 50\,Myr,
and the error normalisation is $\eta = 2$ (i.e., double the fiducial
errors of \textit{Gaia} measurements).
In these cases, the fitted
ages consistently fall short by about $20 \Myr$.
This is a consequence of a degeneracy in the $Z-W$ plane.
The matter density in the Galaxy is reasonably constant with $\approx 100$ pc of the Galactic plane (formally, for our standard model of the Galactic potential, the density at 100 pc is 85\% of the midplane density),
 making the vertical restoring
force close to linear in a star's distance from the midplane, and thus similar to that of a simple harmonic oscillator.
Consequently, each star in our synthetic associations 
has nearly the same period in the $Z$ direction, and thus for a particular current-day distribution of stellar positions and velocities there is a degeneracy between two possible starting states: stars could have
started at similar heights but a wide range of 
velocities, or with a small range of  velocities but a large variation in 
starting heights.
The phase-space distribution of the stellar population
in the $Z-W$ plane is therefore uncorrelated
at multiple distinct epochs,
separated by
a quarter of the vertical oscillation period, which is $\approx 20$\,Myr. 
Our catastrophic failure mode consists of the MCMC walkers settling into
the first of these many degenerate minima
that yields a reasonable fit in other phase-space dimensions. This
failure mode only occurs for errors larger than usual for \textit{Gaia},
because for smaller errors, constraints in the phase-space components that
lie in the Galactic plane are sufficient to break the degeneracy in the 
out-of-plane directions.
Thus, these failures
are not a concern for practical applications, as long as the
relative uncertainties for the majority of stars in question
do not significantly exceed 100\% that of \textit{Gaia}.

\subsection{Stars with realistic initial kinematics}
\label{sec:fed_stars}
In the previous section, all our synthetic associations
had initial conditions that matched our assumptions
(spherical, uncorrelated initial distribution; instantaneous
gravitational unbinding). 
Here we consider a much more realistic initial 
stellar distribution by using stellar positions and 
velocities drawn from a
simulation of star formation.
The simulation we use is the run referred to as case ``GTBJR'' in \citet{onus_numerical_2018}; its initial conditions and physics are
identical to the ``GvsTMJ''  case presented in \cite{federrath_inefficient_2015},
which includes turbulence, magnetic fields and jet feedback,
but with the addition of  radiation feedback as implemented in
\cite{federrath_converging_2017}.
The simulation tracks the collapse of a molecular cloud
in a $2$ pc cube with periodic boundaries.
In the simulation, stars are represented by sink particles, and
for this test we take the positions and velocities of
the sink particles in the final snapshot of the
simulation as the initial
positions and velocities of synthetic stars. As in 
the previous tests, we choose the absolute position 
and velocity of the stars in Galactic coordinates 
such that their current-day central position and 
velocity match that of LCC. We project the stars 
forward through time for $20$\,Myr, ignoring
any gravitational interactions between them, convert their
positions and velocities to astrometric coordinates, and add
random errors with a distribution equal to our fiducial
\textit{Gaia} DR2 median uncertainty ($\eta=1$; see \autoref{sec:synth_gen}). 
We then run \texttt{Chronostar} on
the resulting synthetic data set. 

\begin{figure*}
    \centering
    \includegraphics[width=\linewidth]{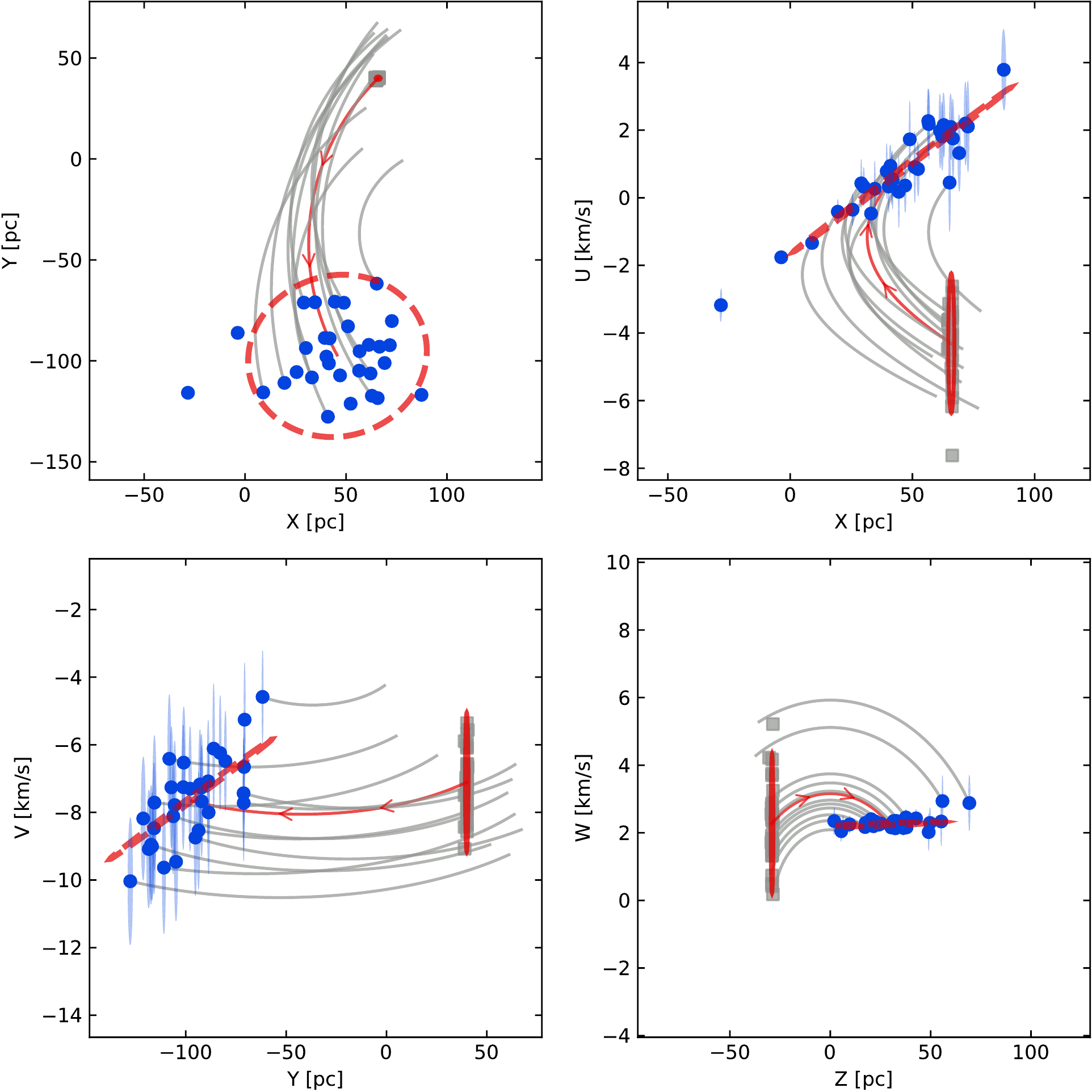}
	\caption{
    Results from our test using synthetic stars drawn from 
    a star formation simulation (\citealt{federrath_inefficient_2015}; 
    \citealt{federrath_converging_2017}).
    Panels show the positions of stars at formation 20\,Myr
    ago (grey squares) and at the current-day (blue circles) in several
    different 2D projections of the 6D phase-space.
    Faint ellipses around the current-day positions show the
    error distributions for the synthetic measurement. The dashed red
    ellipse shows the best-fit current-day distribution retrieved
    by \texttt{Chronostar}; a similar best-fit ellipse for the
    origin point is also plotted in solid red.
    Red lines show the trajectory of the centre
    of the distribution, with arrows pointing forwards through time.
    Grey lines show the trajectories for selected current-day stars
    \textit{traced back} from that
    star's central estimate by 20\,Myr. 
    Note that, due to the relatively
    large uncertainties, these trajectories show minimal
    convergence in $X-U$ and $Y-V$ planes despite being traced back
    to their true seeded age. This is not the case for the 
    trajectories in the $Z-W$ plane, which do indeed show convergence
    in $Z$. This is a consequence of the almost linear restoring force
    in the $Z$ direction, resulting in stars having similar periods. See
    \autoref{sec:synth_gen} for further discussion.
    We note that our choice of a co-rotating reference frame introduces added motion in the $X$ and $Y$ dimensions due to the Coriolis force. As a consequence orbits in the $X-U$ plane curve towards larger values of $X$ despite having $U$ values around $-2 \kms$.
    }
	\label{fig:fed_stars}
\end{figure*}

\texttt{Chronostar} retrieves an age of $20.1\pm0.2$\,Myr,
demonstrating that, in this instance, approximating the initial kinematic
distribution of the association as spherically Gaussian is a
sufficiently accurate approximation. 
\autoref{fig:fed_stars} shows the current and 
starting positions of the stars, along with the fit, 
in 4 different 2D projections ($X-Y$, $X-U$, $Y-V$ and 
$Z-W$). We see that the current-day
fit provides a good match to the current-day position 
and velocity, and the corresponding fit to the origin 
falls within 3 pc of the true average initial position of the stars.
One particularly noteworthy feature of this plot is 
that the correct fit is recovered despite the fact 
that, due to the relatively large uncertainties in 
the current-day kinematic properties of the stars, an 
attempt to trace the stars back in time by 
integrating their orbits does not show any 
significant amount of convergence. Thus attempts to 
reconstruct these stars' origin point by looking for 
a minimum volume or similar, the approach used in 
traceback methods, would be unlikely to succeed.

We also note that the reconstruction strongly favours a single-component fit to this data set. Following the procedure outlined above, after finding a single component fit, \texttt{Chronostar} attempted a two-component fit. However, the BIC of the single-component fit, 546.5, is significantly better than that of the best two-component fit, 576.5.

\subsection{Multiple components}
\label{sec:synth-multi-comps}
Our next tests increase the complexity by introducing
data sets with multiple components, arranged
so that they have significant overlap in position, velocity,
or both with incorporated observational uncertainties 
equal to our fiducial \textit{Gaia} DR2 median uncertainty
($\eta = 1$).
 The goal is to test \texttt{Chronostar}'s ability to
separate such overlapping sets of stars.
We initialise each fit in the default way as described in
\autoref{sec:em}.

In the following text, for convenience we will refer to stars being \textit{assigned}
to components. We remind the reader that \texttt{Chronostar} 
does not assign discrete memberships but rather
utilises continuous, probabilistic memberships. However, 
for convenience of plotting 
and discussion we will describe a star as being assigned to the component for which \texttt{Chronostar} gives the highest membership probability.

\subsubsection{Four distinct components}
\begin{figure*}
	\centering
	\subcaptionbox{\label{fig:four_assocs_xy}}
    	{\includegraphics[width=0.49\linewidth]{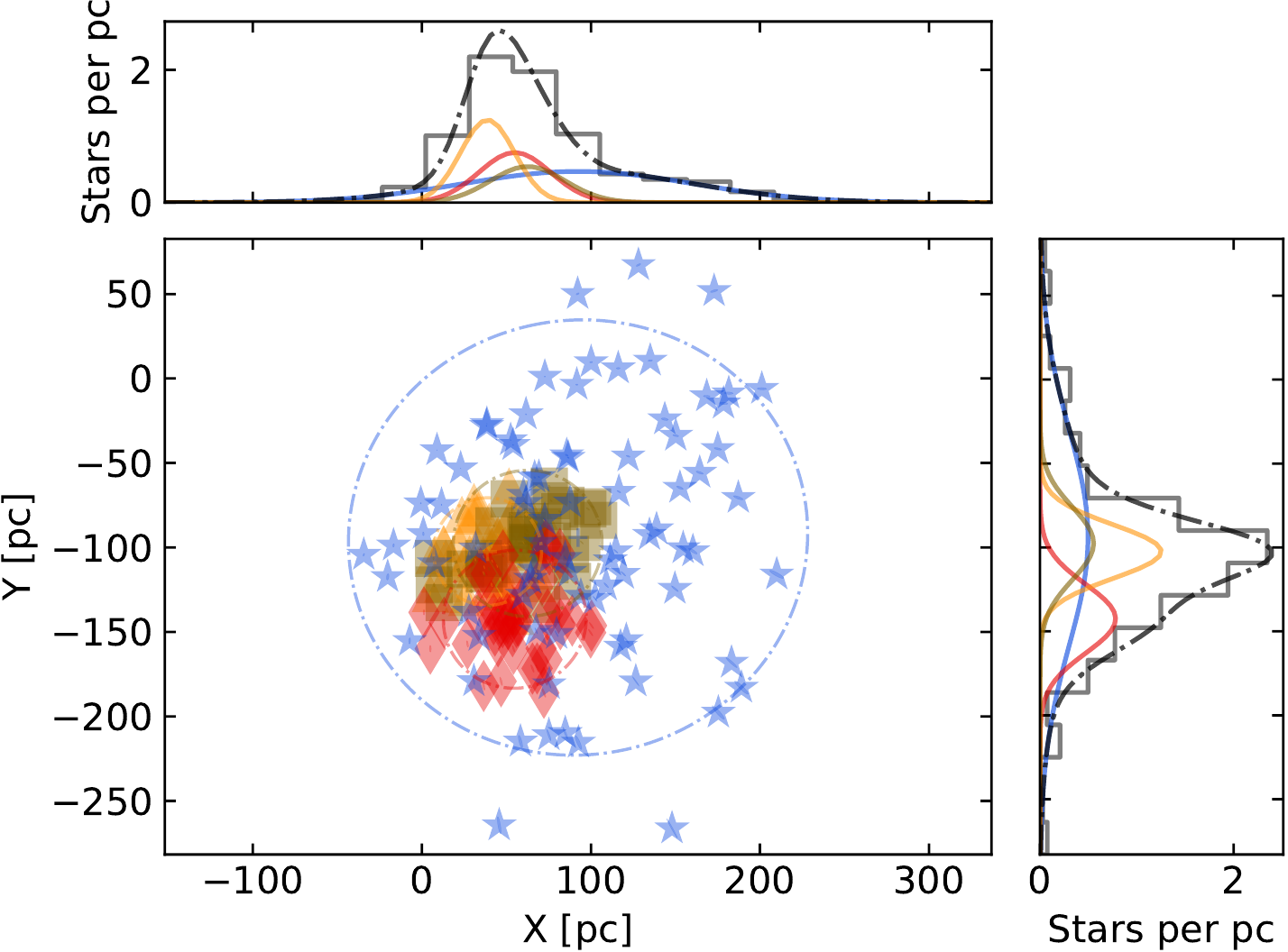}}
	\hfill
	\subcaptionbox{\label{fig:four_assocs_zw}}
      {\includegraphics[width=0.49\linewidth]{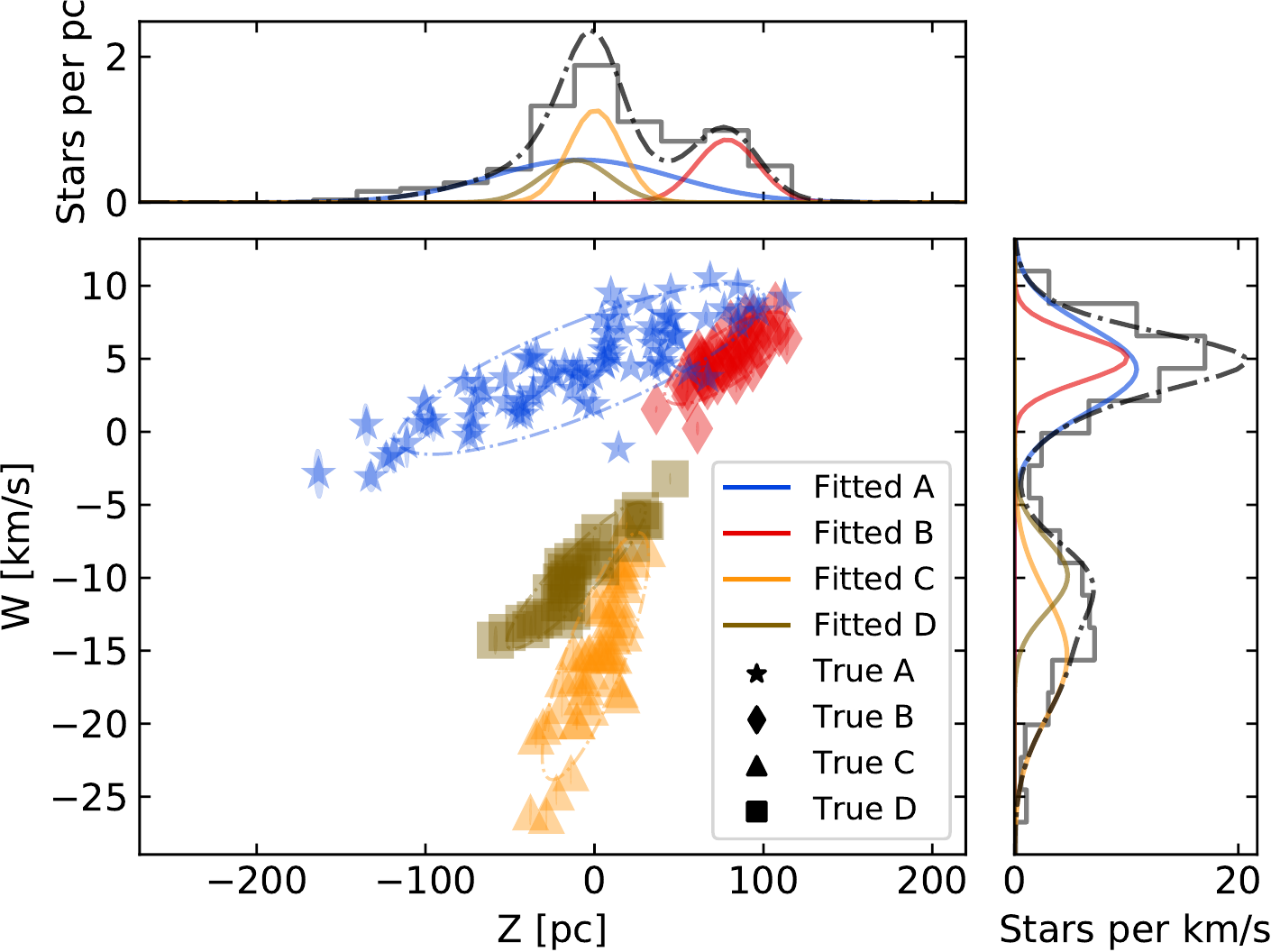}}
	\caption{
    Results from our test using four distinct components with
    distinct ages but overlapping current-day positions.
    Panels show the positions of stars at current-day in several
    different 2D projections of the 6D phase-space. Faint ellipses
    around the current-day positions show the error distributions 
    for the synthetic measurement but are often smaller than the data points.
    The dashed ellipses show the best-fit
    current-day distributions retrieved by \texttt{Chronostar} for each component; see
    \autoref{tab:four-assocs} for the numerical values of the true and fitted parameters of each
    component.
    Marker styles show the component to which each star truly 
    belongs, while marker colours show the component to which 
    \texttt{Chronostar} assigns that star; colours and symbols 
    match in all cases because in this test \texttt{Chronostar} 
    assigns all stars correctly.
    Flanking plots show the 1D projection of densities in each 
    dimension. In these plots, histograms show the distribution of 
    synthetic stars, while curves show the probability densities 
    for \texttt{Chronostar}'s best fit, which colour indicating 
    component. The grey histogram and black line show the sum of 
    all components. 
  }
	\label{fig:four_assocs}
\end{figure*}

The first test 
features four components, each containing 30 - 80 stars, that have
distinct ages from 3 to 13\,Myr yet have current-day distributions that
overlap when viewed solely in position space or solely in
velocity space. However, because these components have 
different ages, they are separable in joint position-velocity 
space (e.g., in the $Z-W$ plane). We give the full set of 
initial parameters for each component, and the best fits to 
them that \texttt{Chronostar} retrieves, in 
\autoref{tab:four-assocs}.  We also show the current-day 
positions of the stars, and
\texttt{Chronostar}'s fits to them, in two 2D projections
of 6D phase-space in \autoref{fig:four_assocs}.

\texttt{Chronostar} successfully fits the ages, initial positions, and dispersions of each component, and correctly
classifies the memberships of all 200 stars, despite the fact that the four components overlap in multiple dimensions. The reason it is able to accomplish this separation becomes clear if we examine panel (b) of \autoref{fig:four_assocs}, which shows the distribution of stars, and our fits to them, in the $Z-W$ plane. Consider the ellipses in (b) in order of ascending age (C, D, B, A):
the angle that the semi-major axes of each component 
makes with the vertical increases systematically with age. In the $Z-W$ plane these angles rotate with time, completing a full rotation after 80\,Myr. Similar rotation occurs with time in the $X-U$ and $Y-V$ planes (not shown), but with a different period. \texttt{Chronostar} is able to separate four components, despite their overlap in both position and velocity, because 
position-velocity correlations provide a sensitive measure
of age since expansion.

\begin{table*}
 \centering 
 	\caption{
 	Parameters for our test using four distinct components with 
distinct ages but overlapping current-day positions. 
    For each parameter, the ``True'' value is the input value 
used to construct the synthetic data, while the ``Fit'' value 
is the 50th percentile value for the posterior PDF returned by 
\texttt{Chronostar}, with error bars indicating the ranges from 
the 16th to 50th and 50th to 84th percentiles. The quantity we 
report for the ``Fit'' value of number of stars (nstars) is the sum of 
the membership probabilities for that component in the final, 
converged output of the EM algorithm.
    }
	\label{tab:four-assocs}
 	\begin{tabular}{l|r|r|r|r|r|r|r|r|}
\hline
& \multicolumn{2}{c |}{Component A}& \multicolumn{2}{c |}{Component B}& \multicolumn{2}{c |}{Component C}& \multicolumn{2}{c |}{Component D}
\\
& True & Fit& True & Fit& True & Fit& True & Fit\\
\hline
$x_0$ [pc]  & $274.8$ & $267.2^{+ 7.2}_{- 6.8}$  & $132.8$ & $120.7^{+ 7.0}_{- 6.4}$  & $ 98.0$ & $ 99.9^{+ 3.8}_{- 4.1}$  & $125.5$ & $123.8^{+ 3.3}_{- 3.1}$ \\
$y_0$ [pc]  & $220.4$ & $216.7^{+ 5.2}_{- 5.3}$  & $ 85.0$ & $ 62.2^{+ 9.9}_{- 9.9}$  & $-46.8$ & $-48.0^{+ 3.0}_{- 3.1}$  & $-52.2$ & $-54.3^{+ 1.9}_{- 2.0}$ \\
$z_0$ [pc]  & $-55.6$ & $-51.6^{+ 2.4}_{- 2.4}$  & $ 11.4$ & $ 18.0^{+ 3.9}_{- 4.1}$  & $ 45.6$ & $ 48.3^{+ 3.0}_{- 2.9}$  & $ 59.6$ & $ 57.7^{+ 2.6}_{- 2.7}$ \\
$u_0$ [$\kms$] & $-17.3$ & $-17.4^{+ 0.6}_{- 0.6}$  & $-12.5$ & $-11.5^{+ 0.5}_{- 0.5}$  & $-21.8$ & $-22.7^{+ 0.6}_{- 0.6}$  & $-12.4$ & $-11.8^{+ 0.6}_{- 0.6}$ \\
$v_0$ [$\kms$] & $-25.1$ & $-25.1^{+ 0.6}_{- 0.6}$  & $-22.7$ & $-23.3^{+ 0.4}_{- 0.4}$  & $-18.4$ & $-18.6^{+ 0.6}_{- 0.6}$  & $ -8.2$ & $ -7.6^{+ 0.6}_{- 0.6}$ \\
$w_0$ [$\kms$] & $  2.6$ & $  1.8^{+ 0.5}_{- 0.6}$  & $  7.8$ & $  7.6^{+ 0.3}_{- 0.3}$  & $-14.6$ & $-15.0^{+ 0.6}_{- 0.6}$  & $ -9.0$ & $ -8.9^{+ 0.6}_{- 0.6}$ \\
$\sigma_{xyz}$ [pc] & $ 20.0$ & $ 19.4^{+ 1.0}_{- 1.0}$  & $ 10.0$ & $  9.6^{+ 0.8}_{- 0.7}$  & $ 10.0$ & $  9.4^{+ 0.6}_{- 0.5}$  & $  7.0$ & $  7.8^{+ 0.7}_{- 0.6}$ \\
$\sigma_{uvw} [\kms]$ & $  5.0$ & $  4.9^{+ 0.2}_{- 0.2}$  & $  2.0$ & $  2.0^{+ 0.1}_{- 0.1}$  & $  5.0$ & $  4.4^{+ 0.3}_{- 0.2}$  & $  3.0$ & $  3.0^{+ 0.3}_{- 0.2}$ \\
age [Myr]   & $ 13.0$ & $ 12.7^{+ 0.3}_{- 0.3}$  & $ 10.0$ & $  9.0^{+ 0.5}_{- 0.5}$  & $  3.0$ & $  3.0^{+ 0.2}_{- 0.2}$  & $  7.0$ & $  6.9^{+ 0.3}_{- 0.3}$ \\
nstars     &  80 & 80.00 &  40 & 40.00 &  50 & 50.00 &  30 & 30.00  \\
\hline
\end{tabular}

\end{table*}

\subsubsection{Two components with shared trajectory}
\label{sssec:shared_traj}
\begin{figure*}
	\centering
	\subcaptionbox{\label{fig:same_centroid_xy}}
    	{\includegraphics[width=0.49\linewidth]{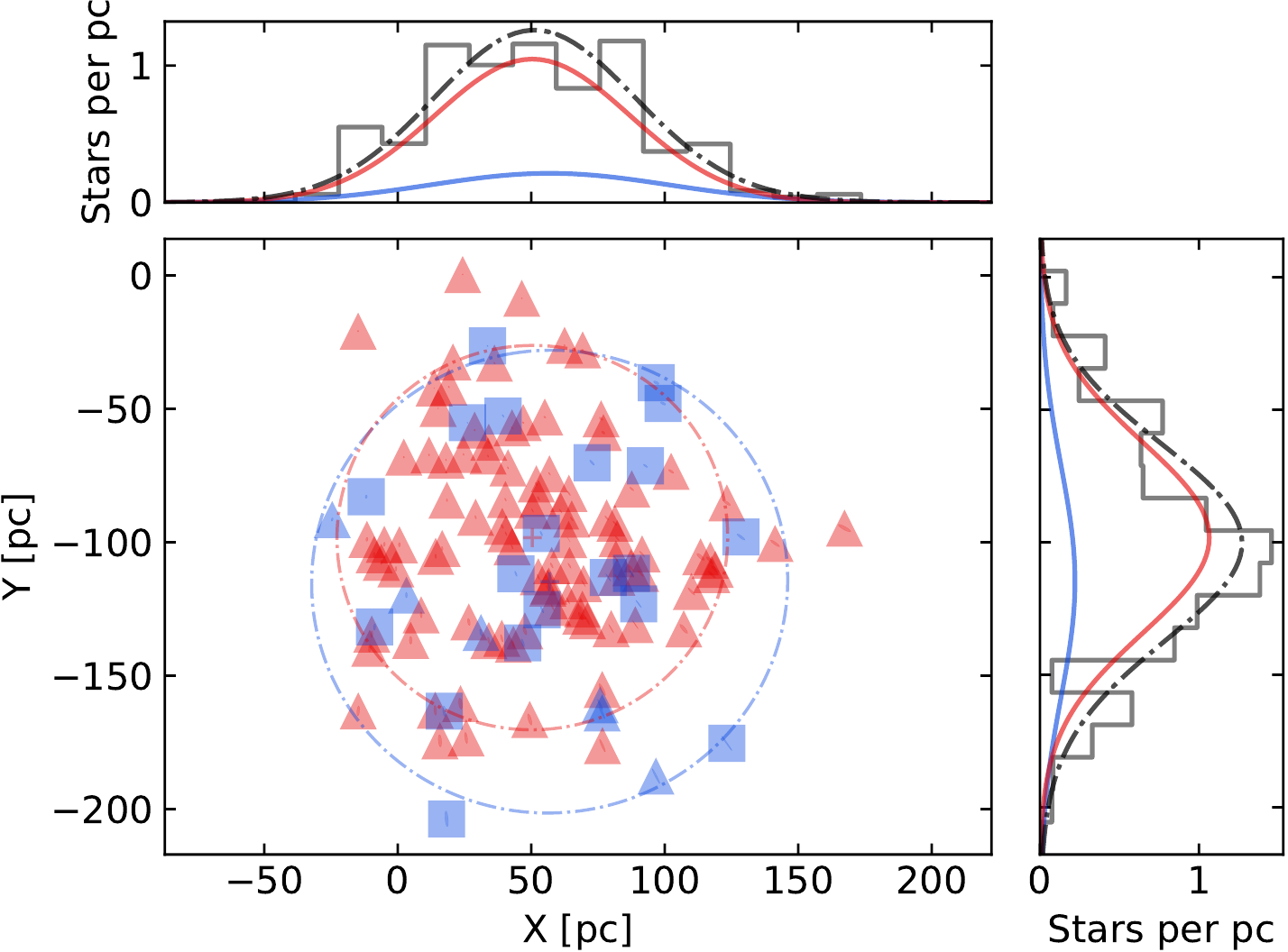}}
	\hfill
	\subcaptionbox{\label{fig:same_centroid_zw}}
      {\includegraphics[width=0.49\linewidth]{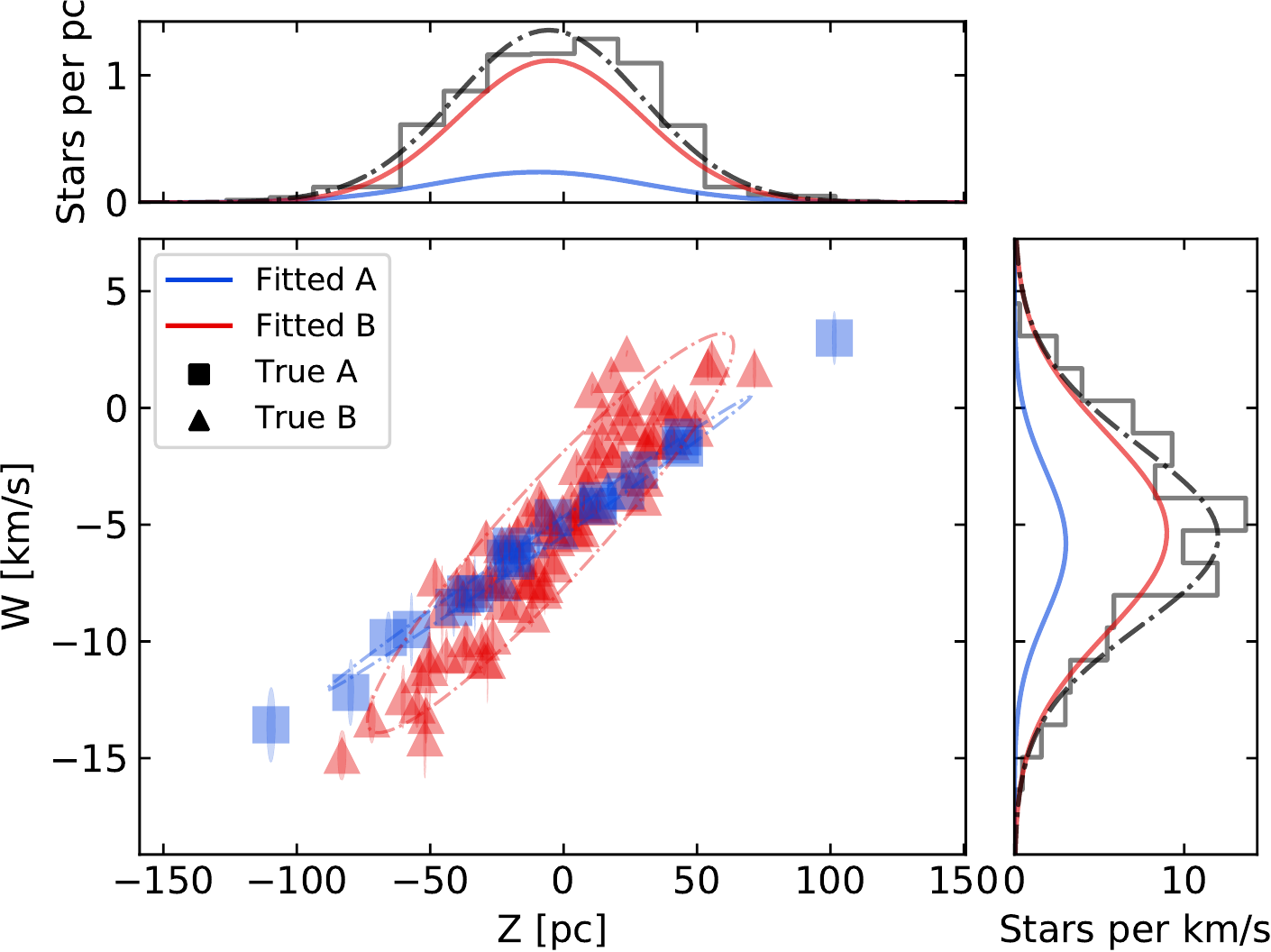}}
	\caption{Results from our test using two components with distinct
  ages but with identical current-day centroids in both position
  and velocity. See caption to \autoref{fig:four_assocs} for an
  explanation of figure details. 
  Note that 6 stars (blue triangles) are misclassified 
  corresponding to a
  success rate of 95\%, which is consistent with the mean membership
  probability that \texttt{Chronostar} estimates for all stars
  with respect
  to their correct component: 94.3\%.
  }
	\label{fig:same_centroid}
\end{figure*}

The second test uses two components with distinct
ages of 7 and 10\,Myr, but with origin points carefully selected
such that the centroids of their current-day distributions
are identical. This gives each association
an identical orbital trajectory, which results
in two distributions
that overlap in every possible 2D projection of the 6D
phase-space.
This scenario presents a challenge to our fitting
approach as there is no separation between the components
along any dimension. The only distinction between
the two components is the \textit{tilt} or degree
of correlation in the \textit{mixed-phase} 
(i.e., position-velocity) planes.
We give the full parameters of the two components used
in this test, and the fits to them derived by
\texttt{Chronostar}, in \autoref{tab:same-centroid}, and
we show two projections of phase-space in \autoref{fig:same_centroid}.
\texttt{Chronostar} recovers the ages of both of our 
overlapping components within a $0.2$\,Myr uncertainty.
Only 6 stars (blue triangles in \autoref{fig:same_centroid})
of 120 are misclassified, corresponding to a
success rate of 95\%, which is consistent with the mean
membership probability 
that \texttt{Chronostar} estimates for all stars to their correct
component:
94.3\%. Thus \texttt{Chronostar} not only returns the correct
assignment for the great majority of stars, it provides an accurate
estimate of the confidence level of the assignments as well.

\begin{table}
 \centering 
 	\caption{
 	Parameters for our test using two components with distinct ages but with
 	identical current-day centroids in both position and velocity. See
    caption to \autoref{tab:four-assocs} for an explanation of the meaning of
    various table entries.
    }
	\label{tab:same-centroid}
 	\begin{tabular}{l|r|r|r|r|}
\hline
& \multicolumn{2}{c |}{Component A}& \multicolumn{2}{c |}{Component B}
\\
& True & Fit& True & Fit\\
\hline
$x_0$ [pc]  & $179.7$ & $179.4^{+ 2.6}_{- 2.7}$  & $129.5$ & $128.0^{+ 2.3}_{- 2.3}$ \\
$y_0$ [pc]  & $ 59.1$ & $ 57.7^{+ 0.9}_{- 0.9}$  & $ 19.5$ & $ 16.0^{+ 2.2}_{- 2.1}$ \\
$z_0$ [pc]  & $ 46.1$ & $ 46.2^{+ 0.6}_{- 0.5}$  & $ 34.1$ & $ 31.8^{+ 1.2}_{- 1.3}$ \\
$u_0$ [$\kms$] & $-16.2$ & $-15.8^{+ 0.9}_{- 0.9}$  & $-14.3$ & $-14.2^{+ 0.5}_{- 0.5}$ \\
$v_0$ [$\kms$] & $-16.4$ & $-17.6^{+ 0.9}_{- 0.9}$  & $-17.8$ & $-17.6^{+ 0.5}_{- 0.5}$ \\
$w_0$ [$\kms$] & $ -3.6$ & $ -4.5^{+ 0.9}_{- 0.9}$  & $ -4.3$ & $ -4.6^{+ 0.5}_{- 0.5}$ \\
$\sigma_{xyz}$ [pc] & $  2.0$ & $  1.9^{+ 0.3}_{- 0.2}$  & $ 10.0$ & $ 10.3^{+ 0.5}_{- 0.5}$ \\
$\sigma_{uvw} [\kms]$ & $  5.0$ & $  4.4^{+ 0.4}_{- 0.3}$  & $  5.0$ & $  4.9^{+ 0.2}_{- 0.2}$ \\
age [Myr]   & $ 10.0$ & $  9.9^{+ 0.1}_{- 0.1}$  & $  7.0$ & $  6.8^{+ 0.1}_{- 0.1}$ \\
nstars     &  20 & 23.90 & 100 & 96.10  \\
\hline
\end{tabular}

\end{table}

\subsubsection{Two components against a uniform background}
The third test uses two components with distinct ages of
$12$ and $25\Myr$ and star counts of 50 and 40,
along with a top-hat PDF background
with density 
$10^{-7} [\un{pc}{}\un{km}{}\un{s}{-1}]^{-3}$,
chosen to be representative of the
density of  \textit{Gaia} DR2 stars in the
vicinity of $\beta$PMG (see \autoref{sec:characterise} for 
details). We refer to the two Gaussian components as an 
association, and to the third component as the background.
We set the bounds of the top-hat PDF representing the 
background to be twice the extent of the 
association in each dimension, and centre it on the mid-range of the
association stars. The resulting bounds of the top-hat PDF are:
\begin{align*}
X&= ( -61.4,  197.6) \un{pc}{}, \\
Y&= (-205.5,   12.6) \un{pc}{}, \\
Z&= ( -58.1,   56.8) \un{pc}{}, \\
U&= ( -16.5,   -4.8) \kms, \\
V&= ( -25.4,  -13.9) \kms, \\
W&= (  -9.9,   -0.5) \kms.
\end{align*}
We draw the true kinematics of 810 stars from this distribution as this 
achieves the desired overall density.
It is straightforward for \texttt{Chronostar} to identify the 
association as  it is a prominent over-density in both
position and velocity space. The challenge lies in how
\texttt{Chronostar} handles the membership boundaries 
of an association against a ubiquitous, fixed background
distribution.

\begin{figure*}
	\centering
	\subcaptionbox{\label{fig:synth_bpmg2_xy}}
    	{\includegraphics[width=0.49\linewidth]{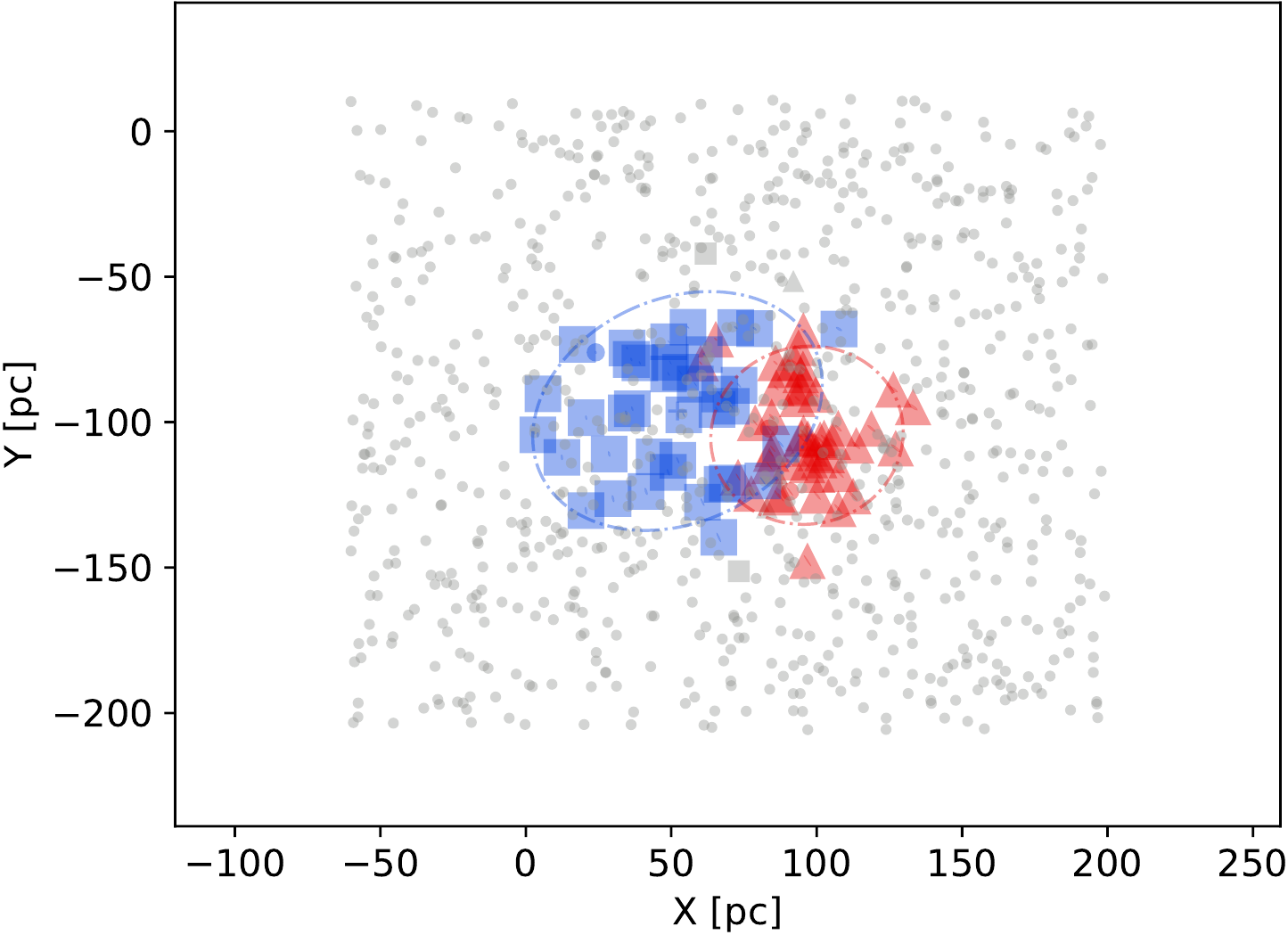}}
	\hfill
	\subcaptionbox{\label{fig:synth_bpmg2_xu}}
    	{\includegraphics[width=0.49\linewidth]{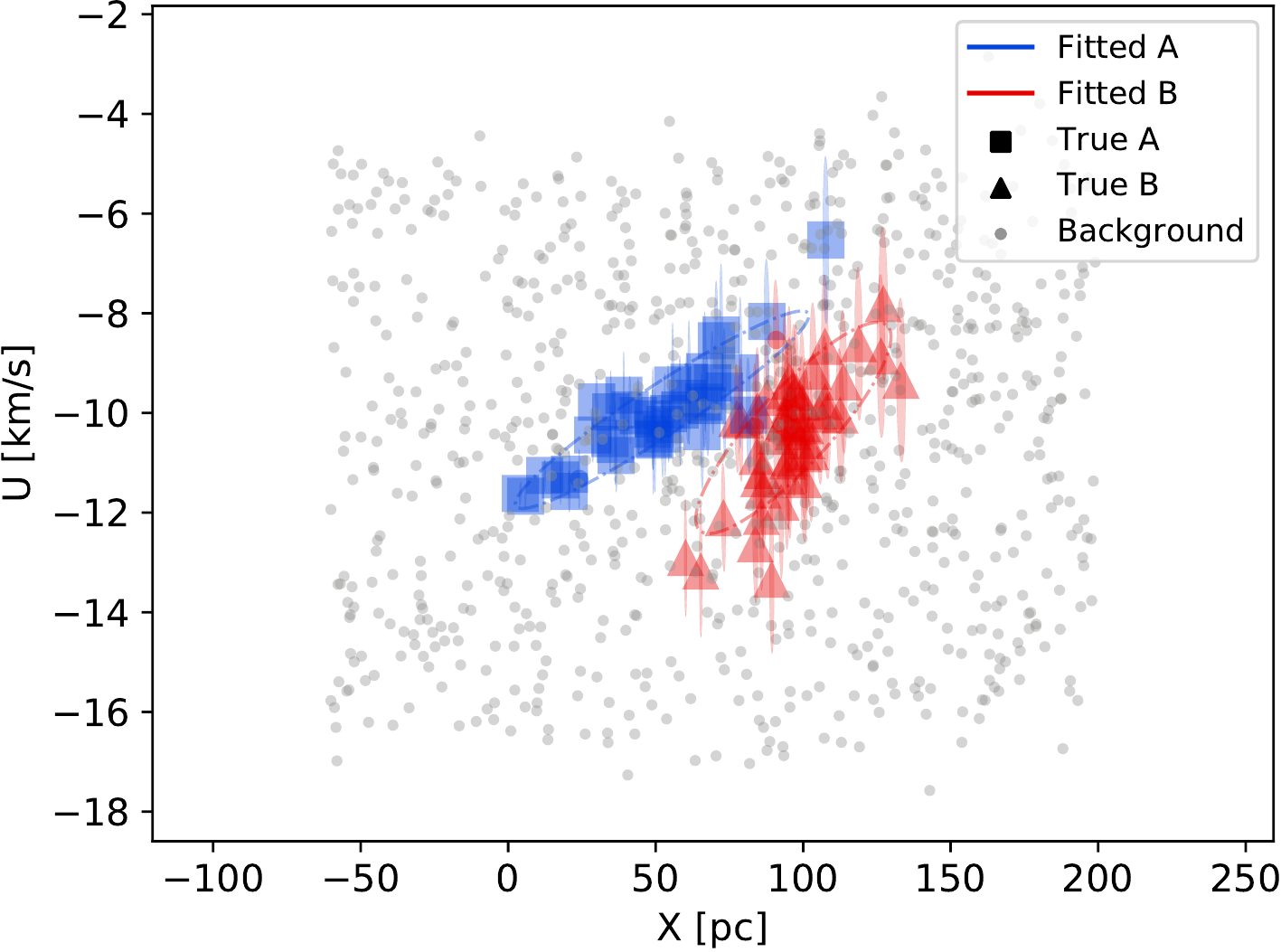}}
	\caption{Results from our test using two components combined with a 
  fixed, uniform background. See caption to \autoref{fig:four_assocs} for
  an explanation for figure details. Grey circles show positions of background
  stars that are correctly classified as part of the 
  background. Grey square and triangular markers show stars that 
  are in fact part of one of the two components of the 
  association, but that are incorrectly classified as part of the 
  background by \texttt{Chronostar}. There are only 5 such 
  markers, because \texttt{Chronostar} correctly classifies
  85 (of 90) component stars.
  }
	\label{fig:synth_bpmg2}
\end{figure*}

We give the full parameters of the two components of the 
association used in this
test, and the fits to them derived by \texttt{Chronostar}, in 
\autoref{tab:synth_bpmg2}, and we show two projections of
phase-space in \autoref{fig:synth_bpmg2}.
\texttt{Chronostar} satisfactorily deduces memberships,
with only 3 (of 810) background stars misclassified as being 
part of the association, and only 5 (of 90)
association stars misclassified as part of the background. Thus the  success rate is
$91.4$\%\footnote{Since the boundary of the uniform background stellar distribution is
arbitrary, and thus the number of background stars is arbitrary,
we disregard correctly assigned background stars when calculating
success rates.}.
This is similar to the mean membership 
probability returned by \texttt{Chronostar} of all component stars
to their true component of origin: 89.2\%.
If we include the membership probabilities of the misclassified
background stars to the background, the mean membership
probability is 86.6\%. In either case, we see both that 
\texttt{Chronostar}'s membership assignments are reasonably 
accurate, and, as importantly, that the level of confidence it 
returns in those assignments is accurate as well.

\begin{table}
 \centering 
 	\caption{Parameters for our test using two components and a uniform
    background. See caption to \autoref{tab:four-assocs} for an explanation
    of the meaning of various table entries.
}
	\label{tab:synth_bpmg2}
    \begin{tabular}{l|r|r|r|r|}
\hline
& \multicolumn{2}{c |}{Component A}& \multicolumn{2}{c |}{Component B}
\\
& True & Fit& True & Fit\\
\hline
$x_0$ [pc]  & $546.5$ & $606.2^{+55.5}_{-58.1}$  & $269.1$ & $278.3^{+24.4}_{-22.8}$ \\
$y_0$ [pc]  & $ 91.5$ & $ 69.2^{+21.9}_{-26.8}$  & $ 63.4$ & $ 64.8^{+ 8.1}_{- 9.1}$ \\
$z_0$ [pc]  & $ 58.7$ & $ 50.9^{+ 5.4}_{- 5.9}$  & $ 51.8$ & $ 52.6^{+ 3.1}_{- 3.3}$ \\
$u_0$ [$\kms$] & $-26.2$ & $-27.5^{+ 1.4}_{- 1.4}$  & $-18.1$ & $-18.7^{+ 0.8}_{- 0.8}$ \\
$v_0$ [$\kms$] & $ -6.2$ & $ -4.4^{+ 1.6}_{- 1.7}$  & $-15.3$ & $-15.1^{+ 0.7}_{- 0.6}$ \\
$w_0$ [$\kms$] & $  2.1$ & $  2.8^{+ 0.6}_{- 0.7}$  & $ -2.9$ & $ -2.7^{+ 0.4}_{- 0.4}$ \\
$\sigma_{xyz}$ [pc] & $ 10.0$ & $ 10.2^{+ 1.0}_{- 0.8}$  & $ 10.0$ & $  9.3^{+ 0.7}_{- 0.6}$ \\
$\sigma_{uvw} [\kms]$ & $  0.7$ & $  0.7^{+ 0.1}_{- 0.0}$  & $  1.0$ & $  1.0^{+ 0.1}_{- 0.1}$ \\
age [Myr]   & $ 25.0$ & $ 27.0^{+ 1.8}_{- 2.0}$  & $ 12.0$ & $ 12.5^{+ 1.1}_{- 1.1}$ \\
nstars     &  40 & 38.22 &  50 & 46.86  \\
\hline
\end{tabular}

\end{table}

\section{Fitting to the $\beta$ Pictoris Moving Group}
\label{sec:bpmg}

In this section we present the results of \texttt{Chronostar}
blindly
applied to 
stars in and around the $\beta$PMG. The goal is to 
recover the majority of $\beta$PMG members, along with a 
viable kinematic age, without any manual intervention.

\subsection{Input data}
\label{sec:input_data}

The first step in our analysis is to prepare a list of stars on 
which to run \texttt{Chronostar}. We start with the list of 
$\beta$PMG members derived by the BANYAN software package 
provided by \citet{gagne_banyan._2018-1}, most of which
are provided with proper motions and radial velocities
compiled from the literature 
(see \autoref{tab:bpmg_members}
for details). We cross-match each 
star in this list with the \textit{Gaia} DR2 catalogue to 
obtain parallax distances along with improved
 proper motions and radial velocities where available.
In cases where a given piece of kinematic 
information is available from multiple sources, we use the 
measurement with the lowest reported uncertainty. We convert 
all astrometric measurements to $XYZUVW$ coordinates as 
described in \autoref{sec:methods-set-up}. 
We then remove stars that lack full 6D phase-space 
information. The result is a set 
of previously-identified $\beta$PMG members with full 6D phase 
space information, including uncertainties. We note that
stars without full 6D phase-space information can be included
by replacing their lacking measurements with placeholder
values with extremely large uncertainties.

We next extend our list by adding stars from the \textit{Gaia} 
DR2 catalogue that have not been identified as $\beta$PMG 
members by BANYAN, but are nonetheless nearby in phase-space. 
To accomplish this, we draw a box in 6D phase-space around the 
centre of the BANYAN stellar list, with its size chosen to be 
twice the span of the BANYAN stars in each dimension. The 
dimensions of this box are $X= (-84.6, 133.1)$ pc,
$Y= (-60.8,  47.3)$\,pc,
$Z= (-34.3,  45.9)$\,pc,
$U= ( -7.1,   9.5) \kms$,
$V= ( -9.4,   2.0) \kms$, and
$W= ( -8.5,   7.5) \kms$. We add to our star list all 
\textit{Gaia} DR2 stars whose central estimates of position and 
velocity fall within this box, and which are not already in the 
BANYAN list. Our final stellar list consists of 859
stars, of which 52 are from 
the BANYAN catalogue and 807 are nearby \textit{Gaia} DR2 
stars that have not been identified as $\beta$PMG members by BANYAN.

Our final data preparation step is to handle binary and 
multiple star systems. The velocities of these stars may 
contain a large contribution from their orbital motion, and 
thus even if the centre of mass velocity of a binary system is 
consistent with being a group member, the component stars may 
be falsely flagged as non-members because their velocities are 
inconsistent with that of the group. To avoid this problem, 
whenever possible we replace multiple star systems with a 
single pseudo-star whose position and velocity (and the 
associated uncertainties on these quantities) are mass-weighted 
averages of the values for the individual stars in the multiple 
system. For stars in the BANYAN catalogue that are flagged as 
multiple, we compute the mass-weighted average by converting 
their spectral types (also taken from the BANYAN catalogue) to 
masses using the conversion table provided by
\citet{kraus_stellar_2007}. Unfortunately we cannot make a 
similar correction for \textit{Gaia} DR2 stars that are not in 
the BANYAN catalogue, because we have no straightforward method 
of identifying which of them are members of multiple systems. 
Consequently, there may be true $\beta$PMG members in the 
\textit{Gaia} DR2 catalogue that are not identified as such by 
\texttt{Chronostar} because their kinematics are contaminated 
by orbital motion. However, since these are only false 
negatives, rather than false positives, we do not expect this 
effect to substantially influence the overall group properties 
that we determine.
This step merges the 52 BANYAN $\beta$PMG members into 38
stars with this change reflected in the following plots.

\subsection{Results of the fit}

\begin{figure*}
    \centering
    \includegraphics[width=\linewidth]{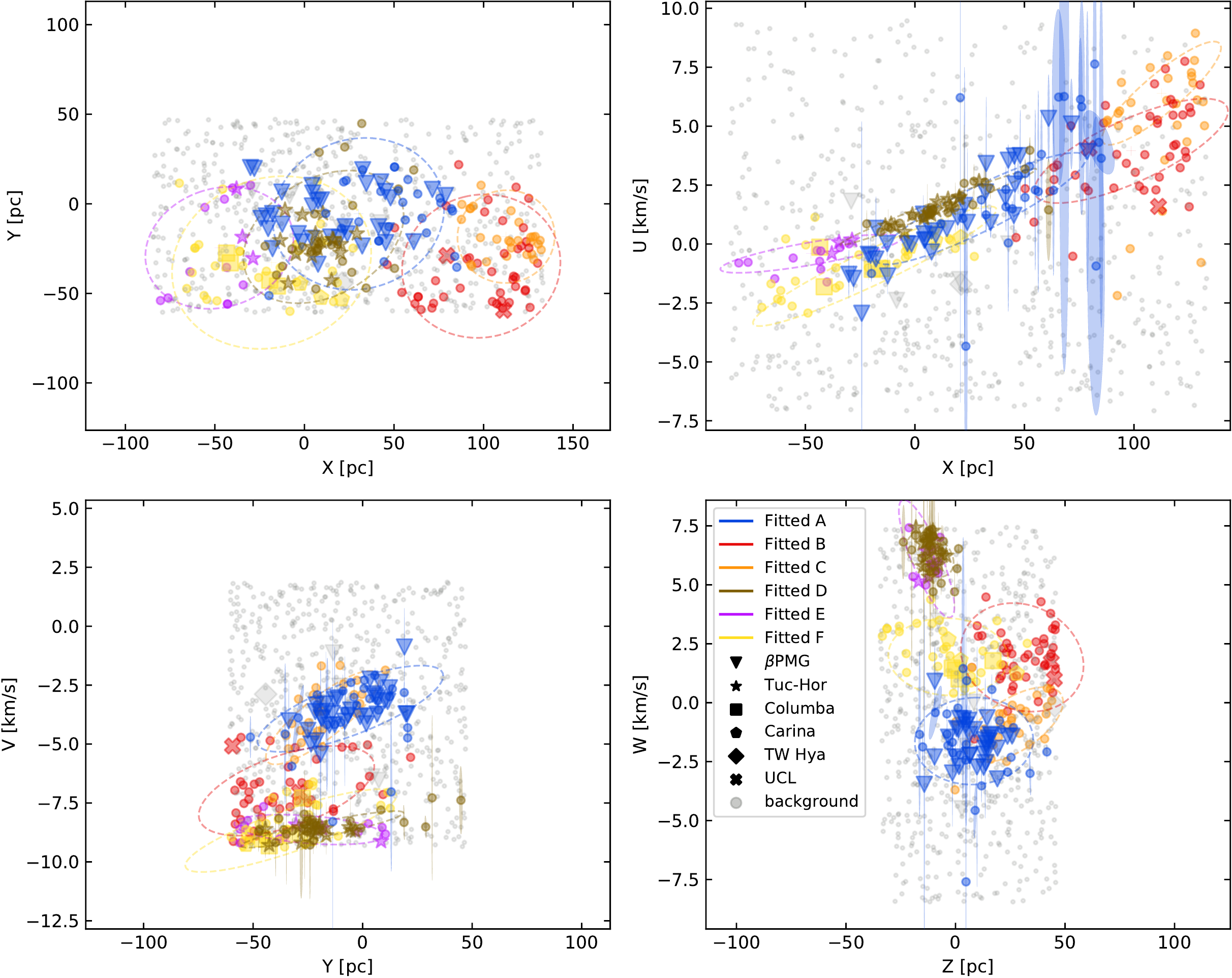}
    \caption{Results from \texttt{Chronostar}'s decomposition
    of stars identified as $\beta$PMG members by BANYAN, plus
    surrounding \textit{Gaia} DR2 stars. Panels show current-day
    positions and velocities of stars in several different
	2D projections. Colours indicate component assignments derived
	by \texttt{Chronostar}, while
	marker styles show assignments of stars to groups and associations 
	in the BANYAN catalog; circles, labelled ``Background'' in 
	the legend, are stars not identified as members of any moving
	group or association by BANYAN, while gray points are
	stars assigned to the background component by \texttt{Chronostar}.
	Filled ellipses around the current-day positions show the error
	distributions for the measurements; for clarity we only
	show error ellipses for probable members of components A and D.
	The dashed ellipses show the best-fit current-day 
    distributions retrieved by \texttt{Chronostar}, 
    coloured by component. 
	Component properties derived by \texttt{Chronostar} are
	provided in \autoref{tab:real-bpmg-table}, while
	a list of likely $\beta$PMG members is given in 
	\autoref{tab:bpmg_members}.
	}
 	\label{fig:bpmg_and_nearby_6E}
\end{figure*}

We run \texttt{Chronostar} on the star list constructed as 
described above. The resulting fit identifies six components, 
which we denote A through F. We show this decomposition in 
\autoref{fig:bpmg_and_nearby_6E}. Of the six components, A 
clearly corresponds to the known $\beta$PMG; it includes 34 of 
the 38 stars in the BANYAN $\beta$PMG catalog, along with an 
additional 27 members that we discuss in detail 
below.\footnote{We remind the reader that 
\texttt{Chronostar} actually returns fractional membership 
probabilities, so when we refer to a star as being identified 
as a member of a particular component, we mean that this is the 
component for which the star has the highest membership 
probability.} The estimated age for this component is 
$17.8\pm 1.2$\,Myr; we report this and other fit results 
in \autoref{tab:real-bpmg-table}. The $\alpha$ value for the
$\beta$PMG component is $50$, indicating that perhaps there
is mass missing in the form of unidentified members.
This component is 
definitively identified as single by \texttt{Chronostar}: a 
six-component model that attempts to divide component A from 
the best five-component into two sub-parts yields a BIC value 
that is 39 higher (see \autoref{tab:bic_table} for summary of BIC values across entire run), strongly favouring the six-component fit.

Of the other five components returned by \texttt{Chronostar}, 
we can identify D with the previously-known Tucana-Horologium 
moving group: of the 46 stars identified as members of this 
component, 17 are listed as Tucana-Horologium members in the 
BANYAN catalogue \citep{gagne_banyan._2018-1}. Our age 
estimate for this component is $36.3^{+1.3}_{-1.4}$\,Myr which is 
consistent with the lithium depletion boundary age estimate of 
$\approx 40$\,Myr given by \citet{kraus_stellar_2014}. We report 
this and other fit parameters for Tuc-Hor in 
\autoref{tab:real-bpmg-table}. However, we warn that, because 
the component that we identify with Tuc-Hor lies at the edge of 
the selection box used to construct our stellar list, and thus 
a substantial number of Tuc-Hor stars are missing from our 
input stellar list, the recovered position and velocity should 
be regarded as unreliable. The slopes in phase-space, and thus 
the age, are more robust to incomplete data. Independent of 
this issue, we emphasise that \texttt{Chronostar}'s 
identification of Tuc-Hor represents a true, blind discovery of 
an association in the \textit{Gaia} DR2 data, since our input 
stellar list was not in any way selected to favour known 
Tuc-Hor stars.

The remaining four components identified by 
\texttt{Chronostar} do not obviously correspond to any known 
associations. Because these components, like Tuc-Hor, lie at 
the edge of our sample selection box, and because unlike 
Tuc-Hor there is at present no evidence that these stellar 
groups are truly coeval, we do not consider their properties 
reliable at this point. We defer further investigation of these 
components to future work.

\subsection{New $\beta$PMG Members}
\begin{figure}
\includegraphics[width=\columnwidth]{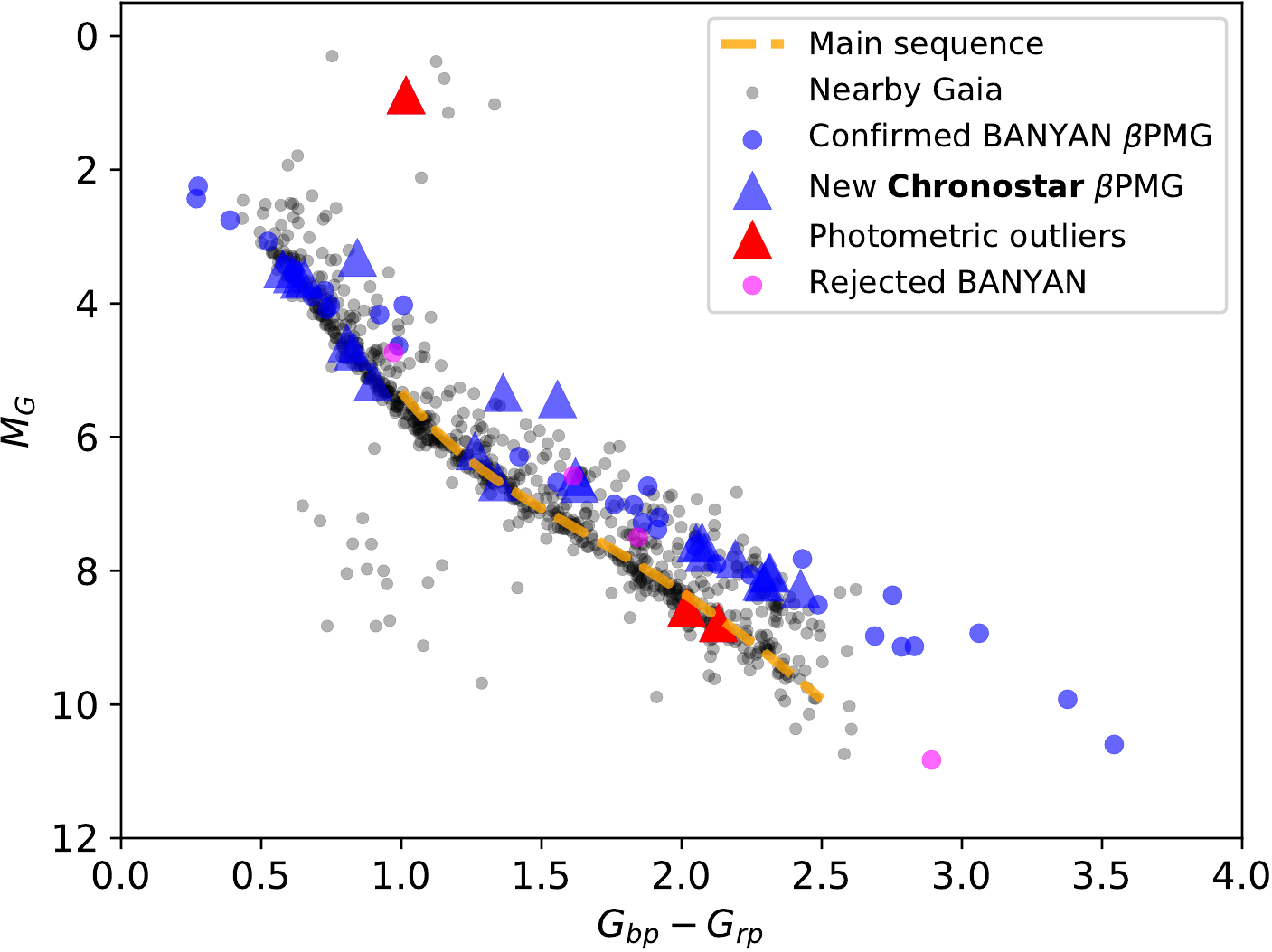}
\caption{
Colour-magnitude diagram of our fit to $\beta$PMG, featuring all
stars in the set described by \autoref{sec:input_data}.
Grey dots are background \textit{Gaia} stars, with a clear
clustering on the main sequence. The dashed orange line
is an empirical fit to the main sequence.
Blue markers denote $\beta$PMG members according to
\texttt{Chronostar}.
Stars that are also $\beta$PMG members according to BANYAN
have circle markers, while those that are new members
have triangle markers. BANYAN members that are rejected by
\texttt{Chronostar} are shown as pink circles.
The three \texttt{Chronostar} members that appear photometrically
inconsistent are coloured red.
}
\label{fig:bpmg_cmd}
\end{figure}

Component A of the \texttt{Chronostar} decomposition, which we 
identify with the $\beta$PMG, has 61 stars with membership 
probabilities greater than 50\% (46 greater than 90\%).  Of 
these, 34 are identified by BANYAN as $\beta$PMG members.
Six further stars are classified as $\beta$PMG members in 
follow up BANYAN papers (\citealt{gagne_banyan._2018}; \citealt{gagne_banyan._2018-2, gagne_volans-carina:_2018}),
and 9 have been identified as likely $\beta$PMG members by 
other authors (see \autoref{tab:real-bpmg-table} for 
references), leaving 13 stars identified by \texttt{Chronostar} 
as likely $\beta$PMG members for the first time. A colour
magnitude diagram (\autoref{fig:bpmg_cmd})
reveals that 10 of these 13 are consistent with lying on an isochrone that is substantially above the main sequence formed by the background stars and consistent with an isochrone formed by previously-identified $\beta$PMG members, further supporting their identification. Thus the Bayesian forward-modelling method 
of \texttt{Chronostar}, coupled with the quality of 
\textit{Gaia} DR2, has allowed us to expand the list of known 
$\beta$PMG members by almost 50\%.

Most of the newly identified $\beta$PMG stars have large $X$ 
and $U$, indicating that the $\beta$PMG extends further in $X$ 
(towards the Galactic Centre) than previously thought.
These stars were likely missed in previous surveys because 
their greater distances imply larger astrometric uncertainties. 
The reason we are able to identify these stars at likely 
members, while previous studies missed them, is that 
\texttt{Chronostar}'s forward-modelling method is significantly 
more robust against uncertainties than earlier traceback methods.

\begin{table}
 \centering 
 	\caption{
 	Parameters for the $\beta$PMG and part of the Tuc-Hor moving 
 	group (see main text) derived by \texttt{Chronostar}'s fit
 	to $\beta$PMG and nearby \textit{Gaia} DR2 stars
 	(Components A and D in \autoref{fig:bpmg_and_nearby_6E},
 	respectively). We suppress
 	here the correlations between dimensions in the covariance
 	matrices, but the full covariance matrices can be found
 	in the online version.
    }
	\label{tab:real-bpmg-table}
 	\begin{tabular}{l|r|r|r|r}
\hline
& \multicolumn{2}{c|}{$\beta$PMG}& \multicolumn{2}{c|}{Partial Tuc-Hor}\\
& Origin & Current& Origin & Current\\
\hline
x [pc]               & $  33.7^{+ 3.8}_{- 3.4}$  & $  30.0^{+ 3.2}_{- 3.1}$  & $ 232.0^{+16.4}_{-16.5}$  & $  12.0^{+ 3.6}_{- 3.6}$ \\
y [pc]               & $  46.6^{+ 3.3}_{- 3.2}$  & $  -5.5^{+ 2.8}_{- 2.8}$  & $ 155.8^{+ 5.3}_{- 6.4}$  & $ -17.2^{+ 3.0}_{- 3.0}$ \\
z [pc]               & $  22.2^{+ 1.9}_{- 1.8}$  & $   7.5^{+ 1.7}_{- 1.7}$  & $ -20.1^{+ 8.7}_{- 8.6}$  & $ -11.8^{+ 0.9}_{- 0.8}$ \\
u [$\kms$]           & $  -0.7^{+ 0.2}_{- 0.2}$  & $   1.5^{+ 0.2}_{- 0.2}$  & $  -7.7^{+ 0.4}_{- 0.4}$  & $   1.6^{+ 0.1}_{- 0.1}$ \\
v [$\kms$]           & $  -3.4^{+ 0.2}_{- 0.2}$  & $  -3.5^{+ 0.1}_{- 0.1}$  & $  -2.5^{+ 0.5}_{- 0.5}$  & $  -8.6^{+ 0.1}_{- 0.1}$ \\
w [$\kms$]           & $   0.2^{+ 0.2}_{- 0.2}$  & $  -1.6^{+ 0.1}_{- 0.1}$  & $  -6.3^{+ 0.2}_{- 0.2}$  & $   6.3^{+ 0.1}_{- 0.1}$ \\
$\sigma_x$ [pc]      & $  12.8^{+ 0.9}_{- 0.8}$  & $  24.5^{+ 1.5}_{- 1.4}$  & $   4.8^{+ 0.4}_{- 0.4}$  & $  23.2^{+ 1.7}_{- 1.5}$ \\
$\sigma_y$ [pc]      & $  12.8^{+ 0.9}_{- 0.8}$  & $  21.6^{+ 1.2}_{- 1.1}$  & $   4.8^{+ 0.4}_{- 0.4}$  & $  19.2^{+ 1.4}_{- 1.2}$ \\
$\sigma_z$ [pc]      & $  12.8^{+ 0.9}_{- 0.8}$  & $  13.7^{+ 0.9}_{- 0.8}$  & $   4.8^{+ 0.4}_{- 0.4}$  & $   5.6^{+ 0.4}_{- 0.4}$ \\
$\sigma_u [\kms]$    & $   1.0^{+ 0.1}_{- 0.1}$  & $   1.2^{+ 0.1}_{- 0.1}$  & $  0.53^{+0.04}_{-0.03}$  & $   0.8^{+ 0.1}_{- 0.1}$ \\
$\sigma_v [\kms]$    & $   1.0^{+ 0.1}_{- 0.1}$  & $   0.9^{+ 0.1}_{- 0.1}$  & $  0.53^{+0.04}_{-0.03}$  & $  0.40^{+0.03}_{-0.02}$ \\
$\sigma_w [\kms]$    & $   1.0^{+ 0.1}_{- 0.1}$  & $   1.0^{+ 0.1}_{- 0.1}$  & $  0.53^{+0.04}_{-0.03}$  & $  0.49^{+0.03}_{-0.03}$ \\
age [Myr]           & -  & $  17.8^{+ 1.2}_{- 1.2}$ & -  & $  36.3^{+ 1.3}_{- 1.4}$ \\
nstars              & -  & $  59.3$ & -  & $  41.1$ \\
\hline
\end{tabular}

\end{table}

\begin{table*}
 \centering 
 	\caption{
 	The BICs scored by various multi-component fits to $\beta$PMG. Each column has an entry for each unique initialisation for the given number of components.
 	In each column the lowest BIC (in bold) is 
 	taken as the best fit for the given component count.
 	The fit that yielded this BIC is then used to initialise multiple
 	fits with $n+1$ components. The row denotes which component
 	from the previous best fit Chronostar decomposes. For example,
 	\texttt{Chronostar} performed two three-component fits,
 	initialising the first by decomposing Component A of the
 	two-component result. \texttt{Chronostar} initialised the
 	second by decomposing Component B, which yielded a better
 	BIC. \texttt{Chronostar} terminated with 6 components because
 	all of the attempted 7 component fits failed to improve the BIC.
    }
	\label{tab:bic_table}
 	\begin{tabular}{l|r|r|r|r|r|r|r}
\hline
& 1 comp & 2 comps  &  3 comps &  4 comps &  5 comps &  6 comps &  7 comps \\
\hline
comp A &\bf{28702.90}&\bf{27984.69}& 28311.91    &\bf{27505.76}& 27566.69    & 27501.87    & 27479.64 \\
comp B &        &             &\bf{27752.75}& 27818.54    &\bf{27449.76}& 27489.28    & 27481.18 \\
comp C &        &             &             & 27761.06    & 27558.12    & 27449.33    & 27506.65 \\
comp D &        &             &             &             & 27563.82    & 27481.90    & 27494.94 \\
comp E &        &             &             &             &             &\bf{27440.57}& 27500.21 \\
comp F &        &             &             &             &             &             &\bf{27448.21}\\
\hline
\end{tabular}

\end{table*}

\begin{table*}
 \centering 
  	\caption{
  	Astrometry and memberships to $\beta$PMG (Comp A)
  	as well as reference to previous work listing star
  	as member of $\beta$PMG.
  	Full table available in machine-readable format in
  	online version.
  	$\beta$PMG membership references: 
(1) \protect\cite{gagne_banyan._2018},
 (2) \protect\cite{schlieder__2010},
 (3) \protect\cite{alonso-floriano_reaching_2015},
 (4) \protect\cite{moor_nearby_2006},
 (5) \protect\cite{elliott_search_2014},
 (6) \protect\cite{moor_unveiling_2013},
 (7) \protect\cite{torres_young_2008},
 (8) \protect\cite{gagne_banyan._2018-1},
 (9) \protect\cite{zuckerman__2001-1},
 (10) \protect\cite{malo_bayesian_2013},
 (11) \protect\cite{malo_banyan._2014},
 (12) \protect\cite{kiss_search_2011},
 (13) \protect\cite{elliott_search_2016},
 (29) \protect\cite{gagne_banyan._2018-2},
 (30) \protect\cite{gagne_pre-gaia_2018},
 (31) \protect\cite{neuhauser_corona_2008}.
 RV references:
 (14) \protect\cite{torres_search_2006},
 (15) \protect\cite{gaia_collaboration_gaia_2018},
 (16) \protect\cite{kiss_search_2011},
 (17) \protect\cite{allers_radial_2016},
 (18) \protect\cite{shkolnik_all-sky_2017},
 (19) \protect\cite{torres_vlba_2009},
 (20) \protect\cite{gontcharov_pulkovo_2006},
 (21) \protect\cite{malo_banyan._2014},
 (22) \protect\cite{valenti_spectroscopic_2005},
 (23) \protect\cite{anderson_xhip:_2012},
 (24) \protect\cite{song_new_2003},
 (25) \protect\cite{montes_late-type_2001},
 (26) \protect\cite{kharchenko_astrophysical_2007},
 (27) \protect\cite{shkolnik_identifying_2012},
 (28) \protect\cite{faherty_population_2016}.
     }
	\label{tab:bpmg_members}
 	\begin{tabular}{l|r|r|r|r|r|r|l|l|l|}
\hline
 Main & R.A. & Decl & Parallax & $\mu_\alpha \cos \delta$ & $\mu_\delta$ & RV & Comp. A & RV ref & Prev.  \\
 Designation & [deg] & [deg] & [mas] & [mas$\un{yr}{-1}$] & [mas$\un{yr}{-1}$] & [$\kms$] & Memb. Prob. &  & $\beta$PMG ref  \\
\hline
HD 203 & $1.709$ & $-23.108$ & $ 25.02\pm0.06$ &$ 96.8\pm0.1$ &$ -47.12\pm0.07$ &$ 7\pm4$ & 0.9836 & 20 & 1  \\
RBS 38 & $4.349$ & $-66.753$ & $ 27.17\pm0.03$ &$ 103.04\pm0.05$ &$ -16.87\pm0.05$ &$ 10.7\pm0.2$ & 0.99927 & 21 & 1  \\
GJ 2006 A & $6.960$ & $-32.552$ & $ 28.66\pm0.07$ &$ 109.8\pm0.1$ &$ -47.39\pm0.07$ &$ 8.8\pm0.2$ & 0.99731 & 21 & 1  \\
| GJ 2006 B & $6.960$ & $-32.557$ & $ 28.68\pm0.08$ &$ 112.2\pm0.1$ &$ -44.64\pm0.07$ &$ 8.5\pm0.2$ & & 21 & 1  \\
Barta 161 12 & $23.808$ & $-7.215$ & $ 26.82\pm0.09$ &$ 97.5\pm0.2$ &$ -49.09\pm0.09$ &$ 6.8\pm0.8$ & 0.93598& 21 & 1  \\
G 271-110 & $24.231$ & $-6.794$ & $ 41.7\pm0.1$ &$ 173.5\pm0.3$ &$ -100.1\pm0.2$ &$ 12.2\pm0.4$ & 0.0 & 27 & 1  \\
HD 14082 A & $34.356$ & $28.745$ & $ 25.13\pm0.04$ &$ 86.92\pm0.07$ &$ -74.07\pm0.07$ &$ 5.4\pm0.5$ & 0.89691 & 22 & 1  \\
| HD 14082 B & $34.353$ & $28.741$ & $ 25.16\pm0.05$ &$ 85.97\pm0.08$ &$ -71.11\pm0.08$ &$ 4.7\pm0.2$ & & 15 & 1  \\
AG Tri A & $36.872$ & $30.973$ & $ 24.36\pm0.05$ &$ 79.68\pm0.08$ &$ -72.00\pm0.07$ &$ 4.8\pm0.1$ & 0.8783 & 15 & 1  \\
| AG Tri B & $36.867$ & $30.978$ & $ 24.45\pm0.08$ &$ 82.7\pm0.1$ &$ -73.49\pm0.09$ &$ 5\pm1$ & & 24 & 1  \\
BD+05 378 & $40.358$ & $5.988$ & $ 22.50\pm0.08$ &$ 79.12\pm0.10$ &$ -56.6\pm0.1$ &$ 5.7\pm0.4$ & 0.0 & 15 & 1  \\
\hline
\end{tabular}

\end{table*}

\section{Discussion} 
\label{sec:discussion}
In this paper we have described the 
\texttt{Chronostar} method for kinematic age 
estimation and membership classification of unbound
stellar associations.
\texttt{Chronostar} models an initial 
association component as a 6D Gaussian with 
uncorrelated positions and velocities, projects this 
forwards in the Galactic potential and maximizes the 
likelihood of the component parameters by overlap 
with current-day stellar measurements. Multiple 
components are treated with an expectation 
maximization (EM) algorithm, and individual components 
have a physical, virial prior on position and 
velocity dispersions. This approach differs from 
\cite{rizzuto_multidimensional_2011} and \textsc{BANYAN} 
\citep{gagne_banyan._2018-1}, which do not 
consider time evolution, and differs from 
\textsc{LACEwING} \citep{riedel_lacewing:_2017}
and \cite{miret-roig_dynamical_2018}, which trace
stellar measurements backwards through time. 

The first distinguishing feature of \texttt{Chronostar} 
is how it handles kinematic fitting and membership assignment
in a self-consistent way, treating the two aspects as a single
problem, and iterating through their circular dependency 
until convergence. Other approaches (e.g. \textsc{BANYAN},
\textsc{LACEwING})
derive association parameters from pre-defined 
membership lists, which in effect (after potential
removal of suspected interlopers) restricts the discovery
of new members to the vicinity of known members.
This also impedes applying constraints on the current
day distributions of associations based on what is
physically plausible.
For example, the classical
decomposition of the Scorpius Centaurus OB association 
into three sub-groups has minimal physical justification
\citep{rizzuto_multidimensional_2011}, and indeed
impedes kinematic ageing techniques when performed on
the large-scale structure enforced by this classification
\citep{wright_kinematics_2018}.

The second significant difference is \texttt{Chronostar}'s
forward modelling of an initial, compact distribution
through the Galactic potential to its 
current-day distribution, intrinsically anchoring
the various variances and covariances of all
dimensions of the 6D ellipsoid to the modelled age.
One obvious benefit of this
approach is the provision of kinematic ages.
A second and less obvious benefit is that the tight 
position-velocity correlations induced by the motions of stars 
through the Galactic potential allow us to more confidently 
reject interlopers that fall well within the extent of the 
distribution of association members in one or more dimensions 
(position or velocity), but do not lie on the correct 
position-velocity correlation. This approach is similar to the 
principle behind expansion ages
(e.g. \citealt{torres_young_2008}, \citealt{wright_kinematics_2018}),
but whereas past applications assume linear expansion, \texttt{Chronostar} accounts for the effects
of the Galactic potential on stellar orbits.
This difference is crucial in pushing to ages $\gtrsim 10-20$ Myr, because the vertical oscillation period of stars through the Galactic plane is $\approx 80$ Myr (74 Myr in our model of the Galactic potential, and $87\pm4$\,Myr
using Oort constants from \citealt{bovy_galactic_2017}).
Position-velocity correlations rotate 90$^\circ$ in the $Z-W$ plane
over a quarter period
(c.f.~the discussion in \autoref{sec:synth_gen}),
so the assumption of purely linear expansion begins to fail
seriously after only $\approx 10-20$ Myr.
A third benefit to forward modeling as done in
\texttt{Chronostar} is that it is 
considerably more robust than traceback or similar methods
against observational uncertainties. Typical radial velocity
errors are $\approx 1$ km s$^{-1}$ 
\citep[e.g.,][]{kraus_spatial_2008}, comparable to the 
intrinsic velocity dispersions of associations. As a result, as 
one attempts to trace stars backward, the volume of possible 
stellar positions balloons rapidly. Attempts to sample this 
volume using Monte Carlo or similar techniques have thus far 
proven relatively unsuccessful at delivering reliable kinematic 
ages \citep[e.g.,][]{donaldson_new_2016, riedel_lacewing:_2017, 
miret-roig_dynamical_2018}. In contrast, trace-forward combined 
with analysis of the overlap between a proposed association and 
observed stars in 6D phase-space does not suffer from this 
explosion of possibilities, because the phase-space volume 
occupied by a proposed stellar distribution is conserved as one 
traces it forward.

\section{Conclusion and Future Work}
\label{sec:conclusion}
In this paper we have presented the methodology for
\texttt{Chronostar}, a new kinematic analysis tool to
identify and age unbound stars that share a common origin.
The tool requires no manual calibration or pre-selection,
and simultaneously and self-consistently solves the problems of 
assigning stars to associations and determining the properties 
of those associations. We test \texttt{Chronostar} extensively
on synthetic data sets, including ones containing multiple,
overlapping components and ones where the initial positions
and velocities are stars are drawn directly from a hydrodynamic
simulation of star formation, and show that it returns very
accurate membership assignments and kinematic ages. In
tests on real data, we show that \texttt{Chronostar} is
capable of blindly recovering the $\beta$ Pictoris and Tucana-Horologium moving groups
(\autoref{fig:bpmg_and_nearby_6E}),
with the kinematic data used to find the latter originating
solely from \textit{Gaia} DR2.
In the future we intend to apply \texttt{Chronostar} 
 to other known associations,
with incorporated radial velocities from dedicated
spectroscopic surveys (i.e. RAVE \citep{kunder_radial_2017}, GALAH 
\citep{buder_galah_2018-1}).
This should for the first time provide reliable kinematic
ages. Since \texttt{Chronostar} has proven to be capable of 
blind discovery, we also intend to search the phase-space near 
the Sun for previously-unknown associations.

Due to the Bayesian nature of \texttt{Chronostar} it is
also straightforward to 
extend it by adding extra dimensions to the
parameter space for even stronger membership classification.
This includes placing priors on the ages of individual stars based on spectroscopic types,
and incorporating chemical tagging into the fitting mechanism.
Two further possible enhancements that we intend to 
pursue in future work include allowing for the 
possibility that associations might be born with 
significant position-velocity correlations (as 
suggested for example by 
\citealt{tobin_kinematics_2009}
and \citealt{offner_stellar_2009-1}),
and allowing 
\texttt{Chronostar} to fit
not only unbound associations but also bound open clusters
that are slowly evaporating.

\section*{Acknowledgements}

MJI, MRK, CF, and M{\v Z} acknowledge support from the Australian Research Council through its \textit{Future Fellowships} and \textit{Discovery Projects} funding schemes, awards FT180100375 (MRK), FT180100495 (CF), FT130100235 (MI), DP150104329 (CF), DP170100603 (CF), DP170102233 (M{\v Z}), DP190101258 (MRK), and from the Australia-Germany Joint Research Cooperation Scheme (UA-DAAD; MRK and CF). MRK and CF acknowledge the assistance of resources and services from the National Computational Infrastructure (NCI), which is supported by the Australian Government. TC acknowledges support from the ERC starting grant No. 679852 `RADFEEDBACK'.




\bibliographystyle{mnras}
\bibliography{paper} 




\appendix
\section{Projecting a Component through Time}
\label{ap:project}
This appendix details our method of taking a 
component's initial distribution in 6D phase-space
$\mathcal{N}(\vt;\mu_0, \bS_0)$
and, using \texttt{galpy}'s orbit calculations \citep{bovy_galpy:_2015},
projecting it forward through time by its modelled
age $t_c$ to its current-day distribution 
$\mathcal{N}(\vt;\mu_c, \bS_c)$.

The technical details of the implementation 
used by \texttt{galpy} \footnote{We used the integrator option \texttt{'odeint'} which utilises \texttt{scipy}'s \texttt{odeint}.}
to calculate orbits
are irrelevant for our method so we abstract the 
\texttt{galpy} orbit calculation as a function
$\vf$ that maps a starting point $\vt_0$ in the 6D phase-space
forward over an arbitrary time to a new point $\vt_f$.
Thus we use $\vf$ to project $\vm_0$ and $\bS_0$ to 
their current-day values,
$\vm_c$ and $\bS_c$.
Acquiring the current-day central value is simply a 
matter of evolving the initial central point forward
by the modelled age. In our abstracting notation this reads:
\begin{equation}
	\vm_c = \vf(\vm_0, t_c).
\end{equation}

To transform the covariance matrix $\bS_0$ it is useful
to think in terms of a coordinate 
transformation between two systems: the initial 
coordinates, and the current-day coordinates.
To linear order, the covariance matrix $\bS_B$ in the new coordinate system $B$ is given by the usual law for error propagation,
\begin{equation}
\bS_B = \bj_A \bS_A \bj_A^T,
\label{eq:error_prop}
\end{equation}
where $\bS_A$ is the covariance matrix in the original coordinates $A$, and $\bj$ is the Jacobian for the mapping between the two coordinate systems; specifically,  if some function $\mathbf{g}(\mathbf{a})$ maps a point $\mathbf{a}$ in coordinate system $A$ to a corresponding point in $B$, then
$\bj_{ij} = \partial g_i/\partial a_j$.

In our case of orbit evolution, the mapping function $\mathbf{g}$ is simply $\vf(\vm_0, t_c)$, and thus the Jacobian evaluated at starting phase-space position $\vm_0$ is
\begin{equation}
\bj_{ij} = \frac{\partial}{\partial \theta_{0,j}} \left(f(\mu_0, t_c)_i\right) = \frac{\partial \theta_{c,i}}{\partial \theta_{0,j}},
\end{equation}
where $\theta_{0,i}$ and $\theta_{c,i}$ are the $i$th components of $\vm_0$ and $\vm_c = \vf(\vm_0,t_c)$, respectively. That is, $\theta_{0,i}$ and $\theta_{c,i}$ are the $i$th components of the phase-space coordinates of the association's centre at the time of its birth and at the current time, respectively. We cannot evaluate the Jacobian $\bj$ analytically 
because evaluation of $\vf$ requires numerical integration 
of an orbit through the Galactic potential. Instead,
we evaluate the partial derivatives numerically using 
a second-order approximation:
\begin{equation}
\frac{\partial \tht{c}{i}}{\partial \tht{0}{j}} \approx
\frac{f(\vt_0 + \vh, t)_i - f(\vt_0 - \vh,t)_i}{2|h|},
\label{eq:numeric_deriv}
\end{equation}
where $\vh$ is a six-dimensional vector with 
$h_j = 10^{-5}$ (pc or $\kms$, depending on dimension) and all other components zero.

One might worry about the numerical stability of this procedure, and this would indeed be a concern if the potential through which we were integrating the orbits were tabulated numerically, or contained significant small-scale structure. However, our potential is both analytic and very smooth, and as a result both the orbit integration and the numerical derivative returned by \autoref{eq:numeric_deriv} are extremely robust to changes in $\vh$ as long as all components of $\vh$ are much smaller than the size scale on which the potential varies, and much larger than the $\approx 10^{-8}$ error tolerance in the numerical integrator. To confirm this directly, we have experimented with varying $h$ from $10^{-1}$ to $10^{-6}$. We find that, over this range, the non-zero components of $\partial\theta_{c,i}/\partial\theta_{0,j}$ vary by $<0.01\%$ in the vast majority of cases, and by $<1\%$ in all cases.

\section{Exploring dependence on potential model}
\label{ap:dependence}

The choice of the potential model influences the calculated orbits and thus the resulting age fits. To explore the dependence we repeat the test described in
\autoref{sssec:shared_traj} using two different potential models.
In both cases we use the same synethetic data which we generated using the potential model \texttt{MWPotential2014}, but then use \texttt{Chronostar} to fit the stellar population using a different potential, in order to determine how sensitive our results are to uncertainties in the potential.

The two key timescales of our potential model are the vertical oscillation frequency and the epicyclic frequency. We expect the vertical oscillation period to have the greater effect on on the age fit as this is the shorter timescale by a factor of 2. Therefore in the two following tests we alter the potential by varying the scale height of the disc component. In the first test we halve the scale height, and in the second we double it. These new potentials have vertical oscillation periods scaled by 0.73 and 1.33 respectively.

 Fitting the synthetic data described in \autoref{sssec:shared_traj} using a potential with a halved scale height continues to yield two components, with estimated ages of $6.8^{+0.2}_{-0.1}\Myr$ and $9.9^{+0.1}_{-0.2}\Myr$. The corresponding results for a potential with a doubled scale height are $6.8\pm0.1\Myr$ and $9.9\pm0.1\Myr$. These compare well with the true ages of 7 and 10 Myr, and with the ages determined using the correct potential. Thus despite the vertical oscillation frequency differing significantly in these two tests, and thus distorting the best fit in the $Z-W$ plane, there is sufficient constraining information in the remaining dimensions to accurately retrieve the ages. The implication of this test is that our age estimates are relatively robust to errors in any particular dimension of the potential.


\bsp	
\label{lastpage}
\end{document}